\documentclass[fleqn,usenatbib]{mnras}

\usepackage[T1]{fontenc}

\DeclareRobustCommand{\VAN}[3]{#2}
\let\VANthebibliography\thebibliography
\def\thebibliography{\DeclareRobustCommand{\VAN}[3]{##3}\VANthebibliography}

\usepackage{xspace} 

\usepackage{graphicx}	
\usepackage{amsmath}	
\usepackage{amssymb}	

\newcommand{\sage}{\texttt{SAGE}\xspace}
\newcommand{\dustysage}{\texttt{Dusty~SAGE}\xspace} 
\newcommand{\mentari}{\texttt{mentari}\xspace}

\newcommand{\Msun}{\ensuremath{\mathrm{M}_{\odot}}\xspace}
\newcommand{\Zsun}{\ensuremath{\mathrm{Z}_{\odot}}\xspace}
\newcommand{\Lsun}{\ensuremath{\mathrm{L}_{\odot}}\xspace}

\usepackage{accents} 

\title[Dust contribution to the panchromatic galaxy SED]{Dust contribution to the panchromatic galaxy emission}

\author[D. P. Triani et al.]{
Dian P. Triani,$^{1,2,3,4}$\thanks{E-mail: dian.triani@cfa.harvard.edu}
Darren J. Croton,$^{3,4}$
Manodeep Sinha,$^{3,4}$
Edward N. Taylor,$^{3}$
\newauthor
Camilla Pacifici,$^{6}$
Eli Dwek$^{5}$\\
$^{1}$Center for Astrophysics Harvard \& Smithshonian, 60 Garden Street, Cambridge, MA 02138, USA\\
$^{2}$Research School of Astronomy and Astrophysics, Australian National University, Weston Creek, ACT 2611, Australia\\
$^{3}$Centre for Astrophysics \& Supercomputing, Swinburne University of Technology, Hawthorn, VIC 3122, Australia\\
$^{4}$ARC Centre of Excellence for All Sky Astrophysics in 3 Dimensions (ASTRO 3D)\\
$^{5}$Observational Cosmology Lab, NASA Goddard Space Flight Center, Code 665, Greenbelt, MD 20771, USA\\
$^{6}$Space Telescope Science Institute, Baltimore, MD 21218, USA\\
}

\date{Accepted XXX. Received YYY; in original form ZZZ}

\pubyear{2021}

\begin{document}
\label{firstpage}
\pagerange{\pageref{firstpage}--\pageref{lastpage}}
\maketitle

\begin{abstract}
We have developed a pipeline called \mentari to generate the far-ultraviolet to far-infrared spectral energy distribution (SED) of galaxies from the \dustysage semi-analytic galaxy formation model (SAM). \dustysage incorporates dust-related processes directly on top of the basic ingredients of galaxy formation like gas infall, cooling, star formation, feedback, and mergers. We derive a physically motivated attenuation model from the computed dust properties in \dustysage, so each galaxy has a self-consistent set of attenuation parameters based on the complicated dust physics that occurred across the galaxy's assembly history. Then, we explore several dust emission templates to produce infrared spectra. Our results show that a physically-motivated attenuation model is better for obtaining a consistent multi-wavelength description of galaxy formation and evolution, compared to using a constant attenuation. We compare our predictions with a compilation of observations and find that the fiducial model is in reasonable agreement with: (i) the observed $z=0$ luminosity functions from the far-ultraviolet to far-infrared simultaneously, and hence (ii) the local cosmic SED in the same range, (iii) the rest-frame K-band luminosity function across $0 < z < 3$, and (iv) the rest-frame far-ultraviolet luminosity function across $0 < z < 1$. Our model underproduces the far-ultraviolet emission at $z=2$ and $z=3$, which can be improved by altering the AGN feedback and dust processes in \dustysage. However, this combination thus worses the agreement at $z=0$, which suggests that more detailed treatment of such processes is required.

\end{abstract}




\section{Introduction}
\label{sec:intro}

The state of a galaxy during its evolution depends on various non-linear processes between its components: gas, stars, black holes, dust and metals. This evolution starts with pristine gas falling into the dark matter halo, then cooling to form a disk. Star formation occurs in the galactic disk, polluting the disk with metals and dust. Massive stars evolve into supernovae and reheat the surrounding gas. This heating suppresses future star formation and is known as feedback. Energy from the accretion disk around a supermassive black hole also acts as a source of feedback that can prevent future star formation.  

Dust makes a vital contribution to the conditions for star formation: it acts as an efficient gas coolant \citep{OS73, Dwek87}, provides surfaces for the formation of molecular hydrogen, and shields the molecules from ionizing radiation \citep{HMK79}. Therefore, dust and star formation in a galaxy are often correlated. A description of stellar populations, dust and gas in a galaxy can be inferred from its far-ultraviolet to far-infrared (panchromatic) emission. The spectral energy distribution (SED) describes the intensity of this emission as a function of wavelength. A significant fraction of light from a star-forming galaxy is expected to be absorbed by dust and re-emitted in the infrared. Therefore, many star forming galaxies are often firstly discovered in the infrared. This importance of dust is even more substantial at high redshift, where star formation is more active.

Molecular clouds, the site of star formation, are generally dustier than the diffuse interstellar medium. Young stars, which contribute most of the ultraviolet emission, are still embedded in these clouds. Their light is more heavily affected than the light from older stars whose birth clouds have disrupted \citep{CF00}. In the \citet{CF00} model, birth clouds have a higher optical depth than the diffuse ISM to account for the thicker dust. While dust in the diffuse ISM can affect the light from both young and older stars, dust in birth clouds is only important for stars below an age limit, typically 10 Myr.

Dust is an essential element that shapes the galactic SED across the panchromatic window. An attenuation curve describes how dust absorbs and scatters starlight in the far-ultraviolet to optical wavelengths. In general, the attenuation decreases with increasing wavelength. A popular attenuation curve that is often adopted is an empirical curve obtained from starburst galaxies \citep{Calzetti00}. However, observations of our Galaxy and neighbouring galaxies show variation in the extinction curves \citep{Gordon03}, which is expected due to different dust properties of each galaxy. Therefore, we need to be careful when adopting a fixed attenuation curve (such as the Calzetti curve) as it might not be the best representation of the dust properties in a galaxy. Another commonly used model is \citet{CF00}, which is theoretically based and allows for varying of the attenuation curve parameters.

The absorbed light is then re-emitted by dust at much longer wavelengths in the infrared region. The shape of the infrared SED depends heavily on the dust grain properties, such as its temperature distribution, grain size and grain composition. There are various infrared spectral libraries, including those derived empirically, theoretically, and both. Empirical spectra often only have one parameter determining the changes \citep[e.g.,][]{Rieke09} while theoretical templates are multi-dimensional \citep{DL07, daCunha08}, and semi-empirical templates sit in between \citep[e.g.,][with two parameters]{Dale14}.

Given the importance of dust in multiple parts of the panchromatic SED, a proper dust treatment is critical for inferring galaxy properties from multi-wavelength observations and constructing synthetic SEDs from theoretical galaxy formation models. SED modelling codes are made to interpret the observations and derive many fundamental galaxy properties, such as stellar mass, star formation rate, gas mass, and dust mass. Such codes typically consist of recipes for the star formation history, stellar population synthesis model, dust attenuation, and dust emission, and use a fitting technique to assign galaxy properties based on the observed photometry. Currently, many SED codes explore different assumptions for both stellar and dust emission. For example, while most SED codes adopt a parametric formula for the star formation history, \citet{Pacifici12, Pacifici15} use stochastic star formation from a semi-analytic model (SAM) and combine it with dust attenuation prescriptions. The SED generation code, \textsc{ProSpect} \citep{Robotham20}, allows the flexible construction of star formation and metallicity histories and provides two options for stellar population templates. Variations in the dust assumption were explored by \textsc{MAGPHYS} \citep{daCunha08}, which includes the \citet{CF00} attenuation model and synthetic multi-dimensional infrared templates \citep{daCunha08, daCunha15}, and Cigale \citep{Burgarella05}, that use the \citet{Calzetti00} attenuation model and adopts the infrared templates of \citet{DL07} and \citet{Dale14}. 

In addition to inferring galaxy properties from observations, it is also important to generate mock SEDs from the theoretical modelling of galaxy formation to completely understand how different physical processes change galaxy emissions. Mock catalogues are also critical when developing an upcoming galaxy survey program and an accurate dust treatment is needed to generate realistic catalogues. Unfortunately, both hydrodynamical and semi-analytic models often treat dust attenuation in an ad-hoc manner as a post-processing step to produce simulated SEDs \citep[e.g.,][]{Henriques15, Vogelsberger19}. Such models usually make predictions only in the optical to near-infrared wavelength range. However, in the past decade, galaxy surveys have broadened their wavelength coverage to the far-ultraviolet to far-infrared. These surveys include Galaxy And Mass Assembly \citep[GAMA,][]{Driver09} for local galaxies and the Cosmic Assembly Near-infrared Deep Extragalactic Legacy Survey \citep[CANDELS,][]{Koekemoer11} for high redshift galaxies. Such surveys, along with the wealth of data obtained in the ultraviolet \citep[GALEX,][]{Martin05}, optical \citep[SDSS, 2dFGRS,][]{2000York, 2001Colless}, near-infrared \citep[2MASS,][]{Skrutskie06}, mid and far-infrared \citep[WISE,][]{Wright10} challenge the theoretical models to provide predictions of galaxy emission across the entire panchromatic window.

Radiative transfer codes have been developed to better model the interplay between stars, gas and dust in producing panchromatic galaxy emission \citep{Silva98, Jonsson10}. These codes require information about the distribution of stars, dust and gas as inputs. \citet{Silva98} used a simplified galaxy model that consists of a bulge and a disk component. Other work has taken the description of galaxies from hydrodynamic simulations \citep{Jonsson06, Lanz14}. Although more accurate, such modelling requires expensive computational resources. When considering cosmological scales and very large numbers of galaxies, some effective approximation needs to be made. 

SAMs offer an alternative method to model galaxy formation and their emission across cosmological scales for a cheaper computational cost. \citet{Baugh05} coupled the \textsc{GALFORM} SAM with a full radiative transfer code, \textsc{GRASIL} \citep{Granato00}. Their model predictions in the far-ultraviolet to near-infrared were in good agreement with observations. But, there was tension in the far-infrared wavebands, where the model underproduced the bright end of the far-infrared luminosity function. \citet{Lacey16} found that this tension persists even with an updated version of \textsc{GALFORM} and a different far-infrared emission model. They suggested changes to starbursts' stellar initial mass function (IMF) as a possible resolution.

The SAM of \citet{Somerville12} used a different approach to produce the far-ultraviolet to far-infrared emission from simulated galaxies. They convolved the predicted star formation and metallicity histories from the SAM with the stellar population synthesis code of \citet{2003BC}, then used the two-component attenuation model of \citet{CF00} and explored several infrared templates to predict dust effects on the SED. \citet{Somerville12} scaled the attenuation parameters with computed physical properties such as the gas-phase metallicity, cold gas mass and disk radius. The scaling was tuned to match the observed ultraviolet luminosity functions at $z=0$, and the empirical relation between $L_\mathrm{IR}/L_\mathrm{UV}$ and the bolometric luminosity; $L_\mathrm{IR}$ is the total luminosity emitted by dust in mid-to far-infrared, and $L_\mathrm{UV}$ is the far-ultraviolet luminosity. The authors were able to reproduce the observed far-ultraviolet to near-infrared luminosity functions but underproduced the number of submillimetre galaxies, independently confirming the tension found in \textsc{GALFORM}.

\citet{Lagos19} followed the approach of \citet{Somerville12} to adopt the \citet{CF00} attenuation model when predicting the panchromatic SED from their \textsc{SHARK} SAM. But instead of tuning the parameters to match the observations, they used a correlation between the parameters and dust surface density from \citet{Trayford20}. This correlation was extracted from the coupling of the \textsc{EAGLE} hydrodynamic simulation and a radiative transfer code. However, because \textsc{SHARK} did not directly track dust mass or dust surface density, they tested several prescriptions to compute dust mass from the gas mass and gas-phase metallicity when deriving the attenuation. The authors then added dust emission from the \citet{Dale14} infrared template. With this, they were able to simultaneously reproduce the observed far-ultraviolet to far-infrared emission from galaxies, without the need to change the IMF.

Predictions for the panchromatic SED from a SAM can be improved by computing the dust mass directly in the model instead of using gas mass and metallicity to proxy dust mass. Several SAMs have incorporated a detailed dust prescription which includes stellar dust production, grain growth, and dust destruction by supernova shocks \citep[e.g.,][]{Popping17, Vijayan19, Triani20}. These models can reproduce the observed dust mass function and dust scaling relations at various redshifts. However, the work of producing SEDs from such models is yet to be made. 

In this paper, we use the \dustysage SAM from \citet{Triani20} to predict the galaxy emission in the far-ultraviolet to far-infrared wavelength range. We adopt the attenuation model of \citet{CF00} and derive the attenuation parameters using several approaches, including utilising the dust mass directly computed by \dustysage. We then explore several infrared templates using different techniques. 

This work is captured in a new pipeline called \mentari to generate the panchromatic SED from \dustysage. \mentari is designed to be user friendly and includes a web app version\footnote{\url{https://share.streamlit.io/dptriani/mentari_web/main}} which allows users to combine spectra from the \textsc{GALAXEV} stellar population library \citep{2003BC}, apply the attenuation model of \citet{CF00} and add infrared emission from \citet{Dale14}. The full version of \mentari\footnote{\url{https://github.com/dptriani/mentari}} has functions to extract galaxy properties from \dustysage directly, provides an additional infrared template from \citet{Safarzadeh16}, and convolves the simulated SED through telescope filters to produce synthetic fluxes or AB magnitudes.

The main goal of this work is to investigate the importance of dust in various wavelengths across the panchromatic windows. We focus on the emission of local galaxies because \dustysage is calibrated only to the properties of galaxies at redshift $z=0$. We explore several prescription of dust attenuation and dust templates and investigate their behaviour in different wavelengths. We also discuss the contribution of stellar and AGN emission in the corresponding wavebands. Although, our approach in modeling AGN emission is far too simplistic and would not be the main theme of this paper. We plan to improve our SED model to include a more realistic AGN model in the near future.

This paper is organized as follows: in Section \ref{sec:dustysage} we summarise the ingredients of the \dustysage SAM, including how we model dust. In Section \ref{sec:mentari} we introduce our new SED generation pipeline, \mentari. Then, we present our predictions for the panchromatic SED and luminosity functions in Section \ref{sec:results}. We discuss the comparison of our various prescriptions in Section \ref{sec:discussion} and provide conclusions in Section \ref{sec:conclusion}.

\section{The \dustysage Semi-Analytic Model}
\label{sec:dustysage}

The \dustysage semi-analytic model has been described in detail in \cite{Triani20, Triani21} and is developed from the \sage model of \cite{Croton06, Croton16}. We refer readers to those papers for further details. The code for \dustysage is open source and publicly available\footnote{\url{https://github.com/dptriani/dusty-sage}}. The model includes tracking for dust mass in addition to the usual processes for galaxy evolution modelling. This section provides a brief overview of these processes. 

\subsection{The Millennium N-body simulation}

In this work, we run \dustysage on the dark matter merger trees constructed from the Millennium N-body simulation \citep{Springel05}. The Millennium simulation follows the cosmological parameters from the first-year WMAP results \citep{Spergel03}. It contains $2160^3$ particles with a mass resolution of $8.6 \times 10^8 h^{-1}$ \Msun within a box of side-length $500 h^{-1}$ Mpc.

The merger tree construction for Millennium were carried out with the \textsc{L-HALOTREE} code \citep{Springel05}. Halos were found using the friends-of-friends procedure \citep{Davis85} and subhalos are identified using the \textsc{SUBFIND} algorithm \citep{2001Springel}. Subhalos are defined to have at least 20 particles.

\subsection{Galaxy formation model}

\dustysage follows the majority of baryonic processes laid out in the \sage model, including (i) baryonic infall; (ii) gas cooling; (iii) reionization heating; (iv) star formation; (v) supernova feedback; (vi) active galactic nuclei (AGN) feedback; (vii) mergers and disk instabilities; and (viii) chemical enrichment; although \dustysage uses updated prescriptions for star formation and chemical enrichment. We summarise each of these processes in the following text.

\dustysage divides baryons in a halo into distinct mass reservoirs. The main gas reservoirs include the cold gas in the galactic disk (the interstellar medium, ISM), hot gas in the halo, and ejected gas, which has been heated and expelled out of the system. Mass is exchanged between these reservoirs following the evolutionary processes of the system. 

\dustysage applies a universal baryon fraction to all simulated halos. The total baryonic mass in the halo depends on this fraction times the halo virial mass. If the virial mass increases, infall gas mass is added to the hot gas reservoir to maintain the baryon fraction. If the virial mass decreases, gas is removed from the ejected gas reservoir (if present), then if the ejected gas has depleted, hot gas is removed.

During the epoch of reionization, gas accretion in low mass systems is suppressed due to the photoionization heating of the intergalactic medium (IGM), to the extent that thermal pressure prevents gravitational collapse onto lower mass subhalos. \dustysage follows the approach presented in \cite{Gnedin00} to reduce the baryon fraction for halos below a filtering mass to account for this suppression. The filtering mass changes with redshift. The parameters for this approach are the redshift when reionization began ($z_0$) and the redshift when the Universe was fully reionized ($z_r$). \dustysage adopts $z_0 = 8$ and $z_r = 7$ from \cite{Kravtsov04}.

In the standard galaxy formation paradigm, the infalling baryons are heated to the virial temperature. These baryons cool down and collapse into the disk due to the halo's angular momentum conservation. To compute the mass of cooling gas at each time step, \dustysage assumes a ``cooling radius'', inside which gas can cool. The cooling radius is defined using the dynamical time of the system and depends on the cooling function from \citet{SD93}. A cooling rate is then calculated by assuming an isothermal density profile for the hot gas.

Cold gas differentiates into atomic and molecular hydrogen. In the galactic disk, star formation is fueled by molecular gas. \dustysage adopts the relation between the surface density of the star formation rate ($\Sigma_\mathrm{SFR}$) and molecular hydrogen ($\Sigma_\mathrm{H_2}$) from \cite{Kennicutt12}:
\begin{equation*}
    \Sigma_\mathrm{SFR} = \epsilon_\mathrm{SF} \Sigma_\mathrm{H_2},
\end{equation*}
where $\epsilon_\mathrm{SF} = 0.005\ \mathrm{Myr}^{-1}$ is the default value for the star formation efficiency.

Stars produce heavy elements that enrich the surrounding ISM. \dustysage adopts the chemical enrichment model of \cite{Arrigoni10} where the contribution of different stellar sources is computed self-consistently. We assume that single stars with mass range $1-8$ \Msun release their metals into the ISM via strong stellar winds when they reach the asymptotic giant branch (AGB) phase which can typically range from 200-400 Myr, depending on the redshift. $40 \%$ of stars with a mass between $3$ and $16$ \Msun are binaries \citep{Francois04} and expel their metals via type Ia supernovae (SN Ia). Massive stars (16 - 100 \Msun) end up as type II supernovae (SN II) and eject their metals in the explosion. 

\dustysage uses a \cite{Chabrier03} initial mass function (IMF) to calculate the mass distribution of AGB stars, SN Ia and SN II at every star formation episode. A grid of stellar yield is applied to each of the stellar sources. We use the C, N and O yields from \citet{Karakas10} for AGB stars; C, O, Mg, Si, S, Ca and Fe yields from \citet{WW95} for SN II stars; and Cr, Fe and NI yields from \citet{Seitenzahl13} for SN Ia stars. 

Energy from supernovae can reheat the surrounding gas and move it to the halo, which drives galactic outflows and suppresses future star formation. In low mass systems with a shallow potential well, supernova feedback can eject all the gas in the disk and halo. The ejected gas is not available for future star formation until it is reaccreted back into the disk. In more massive systems, the AGN is a more effective source for feedback. Following \cite{Croton06, Croton16}, \dustysage applies two modes of AGN feedback: the radio-mode and quasar-mode. In radio-mode feedback, gas accretion onto the supermassive black hole (SMBH) emits energy that heats the halo gas and offsets the cooling process. Quasar-mode feedback is triggered by mergers or disk instabilities with a sudden supply of fresh gas from the galaxy. Rapid gas accretion onto the SMBH results in a quasar that can heat and eject gas from the disk or the entire system. 

Within \dustysage, mergers and disk instabilities move material from the disk to the spheroid component, which offers a means to track morphology. For mergers, if the progenitor mass ratio is above a threshold, a major merger occurs: the disk of both galaxies are disrupted and the stars are placed in the spheroid of the combined system. The disruption is not so severe if the mass ratio is below this threshold, known as a minor merger event. Mergers and disk instabilities can create rapid star formation events known as ``starbursts''. The newly formed stars are placed in the spheroid.

\subsection{Dust mass tracking}

\dustysage includes new prescriptions to explicitly track dust mass as an additional component. Dust related processes mainly occur in the galactic disk, including stellar dust production, grain growth via accretion and grain destruction by supernovae. However, we also include dust destruction via thermal sputtering in the halo and ejected reservoir. We touch on these processes below. More detail is given in \citet{Triani20}.
\begin{itemize}
    \item \textit{Stellar dust production} \\
    Condensation of metals in stellar ejecta is the dominant dust production mechanism in galaxies at early times \citep{Triani20}. The condensation efficiency depends on the environmental condition in each ejecta. While most of the dust formed in AGB winds survives, the reverse shock can destroy newly formed dust in SN II \citep{Micelotta16}. We apply a lower condensation efficiency for SN II dust than the AGB dust to account for these effects. Observations have found an absence of dust in SN Ia ejecta \citep{Gomez12, Dwek16}, likely due to the high velocities of the ejecta and the short timescale of the event. Therefore, we assume that no dust is formed in the SN Ia ejecta.  
    
    \item \textit{Grain growth in molecular clouds} \\
    In high-density molecular clouds, dust grains grow in size and mass by accreting refractory materials as their mantle \citep[e.g.,][]{Jones16, Jones17}. The mantle is loosely bound and prone to destruction. The grain growth rate in \dustysage is adopted from \cite{Popping17} and \citet{Asano13} and depends on the existing dust mass, metallicity and the volume density of the molecular clouds. 
    
    \item \textit{Grain destruction by supernovae} \\
    Supernovae blast waves efficiently cycle dust grains in the ISM back to the gas phase metals. The grain destruction is done primarily via grain-grain collisions and thermal sputtering \citep{DS80, Zhukovska14, Slavin15}. The timescale for such destruction events is described in \cite{DS80, Mckee89} as:
    \begin{equation}
        \tau_\mathrm{destruct} = \frac{M_\mathrm{ISM}}{f_\mathrm{SN} M_\mathrm{swept} R_\mathrm{SN}},
    \end{equation}
    where $M_\mathrm{ISM}$ is the total mass of cold gas in the ISM, $f_\mathrm{SN}$ is the ratio of the destroyed dust to the swept dust mass and describes the efficiency of dust destruction by supernova, $M_\mathrm{swept}$ is the total gas mass swept by a supernova and $R_\mathrm{SN}$ is the supernova rate. 
    
    \item \textit{Dust astration in stars} \\
    The dust content of the ISM is reduced by star formation activity. When stars form in a dust-enriched molecular cloud, they trap the dust inside. Such a process is called astration. The rate for dust astration is proportional to the star formation rate multiplied by the dust-to-gas (DTG) ratio of the molecular cloud. 
    
    \item \textit{Dust in inflows and outflows} \\
    Gas outflows due to feedback carry dust and metals out of the interstellar medium, polluting the halo and ejected reservoirs. When ejected gas is reincorporated back to the system, it also brings its associated dust. This gas flow is assumed to have a similar dust-to-gas (DTG) ratio as its origin (i.e., the gas outflow from the ISM will have the same DTG as the ISM gas, gas in cooling will have the same DTG as the halo). 
    
    \item \textit{Dust destruction in the halo and the ejected reservoir} \\
    Reheated dust in the halo and ejected reservoir is subjected to further destruction by thermal sputtering and grain-grain collisions \citep{Draine79}. This destruction is more efficient with increasing gas density and temperature. \dustysage assumes an isothermal density profile for hot gas in the halo, with temperature at the virial value. The nature of the ejected reservoir is less understood. For simplicity, we assume a uniform density profile and the virial temperature.
    
\end{itemize}

\section{The \mentari SED generator tool}
\label{sec:mentari}

We have developed a new tool, \mentari, to generate galaxy SEDs from \dustysage. In this paper, we focus on simulated galaxies with stellar mass $M_* > 10^8 \Msun$. The generated SED covers the far-ultraviolet to far-infrared wavelength and includes stellar emission, dust attenuation and re-emission in the mid and far-infrared. 

\subsection{Stellar emission}

\mentari uses the stellar population synthesis code of \citet[BC03]{2003BC} to construct the stellar emission in the ultraviolet to near-infrared wavelength range. The BC03 code consists of single-age or simple stellar populations (SSPs) in 221 age grids ranging from 0.1 Myr to 20 Gyr with metallicity values of $Z = 10^{-4}, 4 \times 10^{-4}, 4 \times 10^{-3}, 8 \times 10^{-3}, 0.02$, and $0.05$. We adopt the \cite{Chabrier03} IMF to describe the mass distribution of an SSP in \mentari.

\begin{figure}
    \centering
    \includegraphics[width=0.5\textwidth]{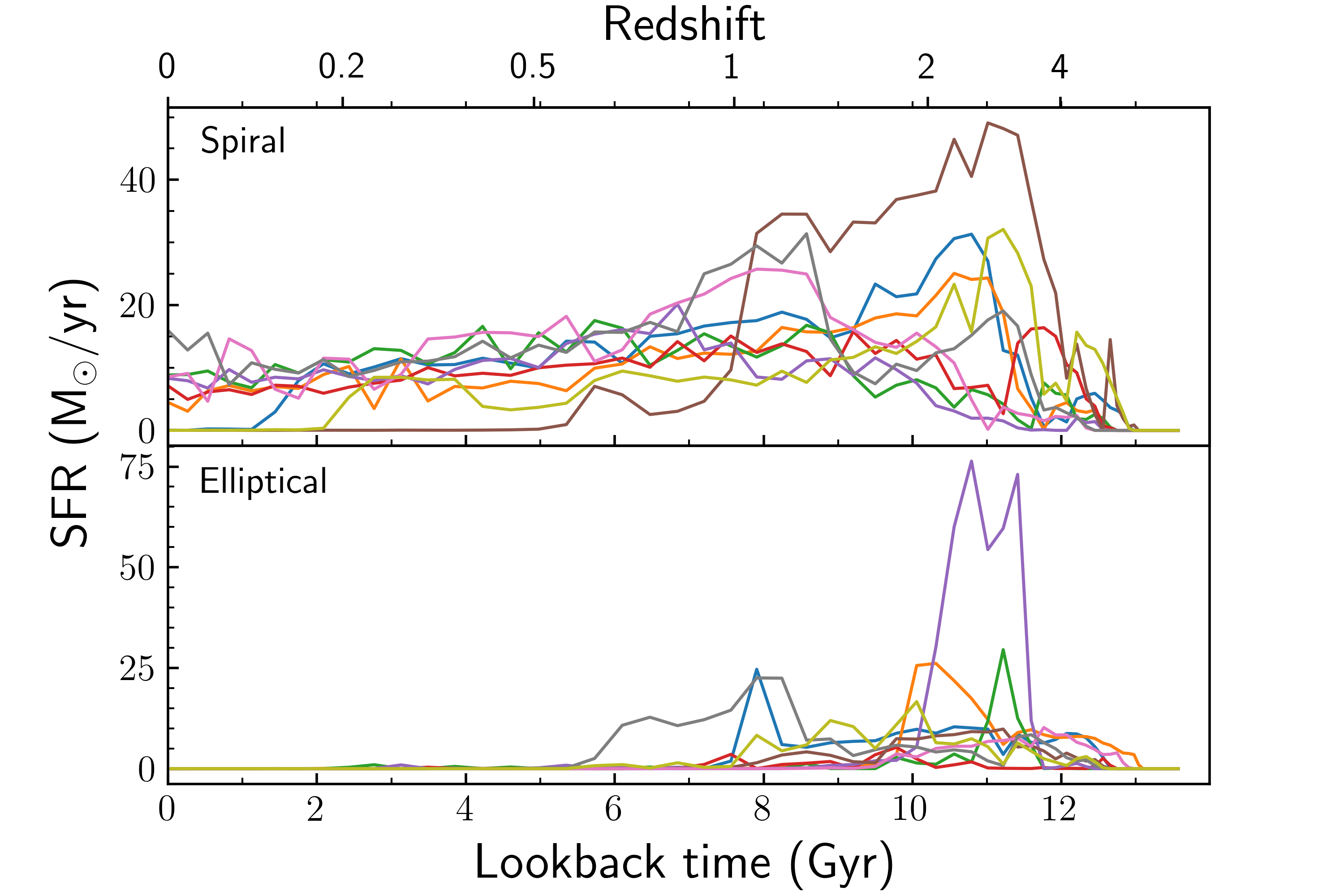}
    \caption{Example star formation rates of spiral galaxies (top panel) and elliptical galaxies (bottom panel) as a function of lookback time extracted from \dustysage. Shown are 10 random galaxies with stellar mass $M_* > 10^{10}$ \Msun for each category. Morphology is defined using a bulge-to-total (BTT) ratio below 0.4 for spirals and above 0.8 for ellipticals.}
    \label{fig:sfh}
\end{figure}

When generating the stellar emission, \mentari accepts star formation and metallicity histories in a tabulated form. This way, it can work with either a parametric or non-parametric star formation histories. Unique for \dustysage, \mentari provides a function to extract the star formation and metallicity histories directly from the simulation output. Figure \ref{fig:sfh} shows examples of the star formation history of spiral galaxies (top panel) and elliptical galaxies (bottom panel) extracted from \dustysage. Morphology is defined using a bulge-to-total (BTT) ratio below 0.4 for spirals and above 0.8 for ellipticals. We select galaxies with stellar mass $M_* > 10^{10}$ \Msun for each category. Most spiral galaxies are still actively forming stars, while elliptical galaxies have lower current star formation rates. Their histories consist of several bursts of various durations and intensities.

\begin{figure}
    \centering
    \includegraphics[width = 0.5\textwidth]{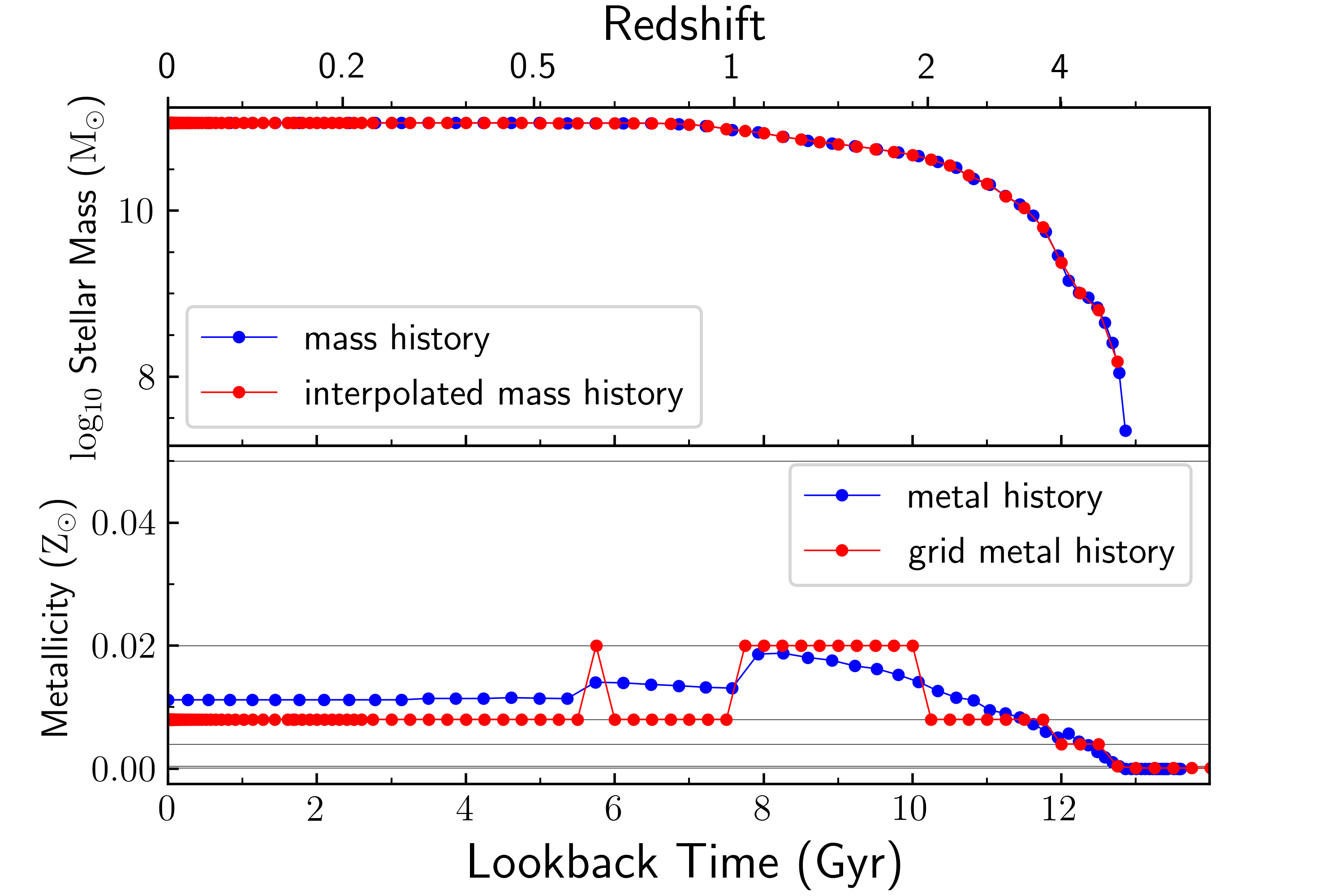}
    \caption{\textit{Top.} Stellar mass of a \dustysage galaxy as a function of lookback time. Blue dots show the stellar mass at 64 time steps from \dustysage, while the red dots are located at the 221 age grids of the BC03 template. \textit{Bottom.} Stellar metallicity of a \dustysage galaxy as a function of lookback time. Grey lines indicate the metallicity grids of the BC03 template, which are $Z = 10^{-4}, 4 \times 10^{-4}, 4 \times 10^{-3}, 8 \times 10^{-3}, 0.02$, and $0.05$. Because of the coarse metallicity grids, we do an interpolation to the metallicity history from the \dustysage timesteps (blue dots) in a similar fashion as in the stellar mass history, then pull the interpolated metallicity to the closest grid point (red dots).}
    \label{fig:interpolation}
\end{figure}

The time grid in the SSP is chosen to be that which most closely maps from the model. However, each BC03 age interval can differ from the timescale between snapshots from the model output. \mentari interpolates the ingested star formation history to the age grid provided by the BC03 template, which is equivalent to rebinning the star formation history onto the BC03 age grid. A star formation history (star formation rate as a function of time, $\mathrm{SFR(t)}$) from \dustysage can be very stochastic, thus interpolating the $\mathrm{SFR(t)}$ can lead to inconsistent total stellar mass across the history. To avoid this, \mentari converts the $\mathrm{SFR(t)}$ into a cumulative stellar mass history and uses it in the interpolation. This approach takes advantage of the narrow time steps in the BC03 age grid so that the chosen SSP in each step adequately describes the stellar population for the duration of the time step.

When constructing the cumulative mass history, the stellar mass formed at a snapshot is computed by multiplying the star formation rate with the time interval to the previous snapshot, corrected by the fraction of mass returned to the ISM by stellar evolution. The total stellar mass at a snapshot is calculated as the sum of masses formed across all snapshots from the initial formation to that particular snapshot. The top panel of Figure \ref{fig:interpolation} shows an example of \mentari interpolation to a smooth stellar mass history; blue dots show the stellar mass at 64 timesteps from \dustysage, while red dots are the 221 age grids of the BC03 template. The bottom panel shows the remapping of the computed metallicity history from \dustysage. The BC03 template has only six metallicity values to choose from, indicated with the grey lines, limiting our ability to model the stellar populations. First, we interpolate the metallicity history from the \dustysage timesteps (blue dots) in a similar fashion as in the stellar mass history. Then, we snap the interpolated metallicity to the closest grid value (red dots). This approach improves the simple assumptions adopted in many SED codes, which either fix the metallicity to a default value (often solar value) or allow the user to vary it but keep it constant across the history. In principle, we could mix the stellar spectra from different metallicities \citep[see e.g.][]{Robotham20}. However, this approach is computationally more complicated and differs from interpolation because metallicity effects at a fixed age are not linear. Therefore, we decided to pick the most appropriate metallicity from the BC03 grid at each timestep, which still allows for metallicity variation across a galaxy's history.

\subsection{Dust attenuation}

Attenuation of the stellar light by dust depends on the star dust geometry and the thickness of dust between the stars and the observer. In galaxies, the light from young stars undergoes heavier attenuation because such stars are still embedded in their molecular clouds with high density dust. \mentari adopts the two-component dust attenuation model of \citet[CF00]{CF00}. The model considers attenuation by the diffuse dust in the ISM and the dust in stellar birth clouds. 

We have tested several methods to compute the attenuation parameters from \dustysage. We expand each of the methods below. Some techniques use the dust surface density as the basis to compute the attenuation parameters. To compute dust surface density from \dustysage, we use:
\begin{equation}
    \Sigma_\mathrm{dust} = \frac{M_\mathrm{dust}}{2 \pi r_\mathrm{0.5,dust} l_\mathrm{0.5,dust}},
    \label{eq:sigma_dust}
\end{equation}
where $M_\mathrm{dust}$ is the total dust mass in the ISM, $r_\mathrm{0.5,dust}$ is the dust half-mass radius, and $l_\mathrm{0.5,dust}$ is the projected minor axis. The dust half-mass radius is defined as $r_\mathrm{0.5,dust} = C r_\mathrm{disk}$. We choose a default value of $C = 0.16$ to match the observed dust surface density profile of SDSS galaxies \citep{Menard10}. The outer disk radius is defined as $r_\mathrm{disk} = 3 \lambda R_\mathrm{vir}/\sqrt{2}$, which is three times the disk scale length \citep{1998MMW} using the Milky Way as a guide \citep{vandenbergh00}. $\lambda$ is the spin parameter of the halo \citep{Bullock01} and $R_\mathrm{vir}$ is the halo virial radius. The minor axis is defined as $l_\mathrm{0.5,dust} = \sin{i} \times (r_\mathrm{0.5,dust} - r_\mathrm{edge-on}) + r_\mathrm{edge-on}$, $i$ is the inclination which we assign randomly to each galaxy, and $r_\mathrm{edge-on} = r_\mathrm{0.5,dust}/7.3$. In computing the edge-on projected radius, the factor $7.3$ comes from the relation of scale-height to scale-length observed in local galaxy disks \citep{Kregel02}.

\subsubsection{Fixed attenuation parameters following \citet{CF00}}
The effective attenuation curve depends on stellar age as the birth clouds dust only affect the light from young stars. The optical depth is given by:
\begin{equation}
    \tau_\lambda (t') = 
    \begin{cases} 
        \tau_\mathrm{ISM} (\lambda/5500 \text{\AA})^{\eta_\mathrm{ISM}} & \text{for}\ t' > t_0,\\
        \tau_\mathrm{ISM} (\lambda/5500 \text{\AA})^{\eta_\mathrm{ISM}} + \tau_\mathrm{BC} (\lambda/5500 \text{\AA})^{\eta_\mathrm{BC}} & \text{for}\ t' \leq t_0,
    \end{cases}
\end{equation}
where $\tau_\mathrm{ISM}$ and $\tau_\mathrm{BC}$ are the optical depths at $5500 \text{\AA}$ for the diffuse ISM and birth cloud components, respectively. $\eta_\mathrm{ISM}$ and $\eta_\mathrm{BC}$ are the powerlaw indexes for both components and $t_0$ is the age threshold for stars embedded in their birth clouds. The default parameters in \cite{CF00} are $\tau_\mathrm{ISM} = 0.3$, $\tau_\mathrm{BC} = 1$, $\eta_\mathrm{ISM} = \eta_\mathrm{BC} = -0.7$.

\subsubsection{Varying attenuation parameters following \citet{Lagos19}}
We follow the method laid out in \cite{Lagos19} to compute the \cite{CF00} attenuation parameters. \cite{Lagos19} computed multiwavelength emission from simulated galaxies in the \textsc{SHARK} SAM \citep{Lagos18} using the \textsc{PROSPECT} SED generator tools. They adopted a parametrization of the \cite{CF00} model from the coupling of the \textsc{EAGLE} hydrodynamical simulation with the \textsc{SKIRT} radiative transfer code \citep{Trayford20}. The diffuse ISM attenuation parameters, $\tau_\mathrm{ISM}$ and $\eta_\mathrm{ISM}$, were found to vary with dust surface density, $\Sigma_\mathrm{dust}$. We use the median and 1-$\sigma$ scaling relations in \citet{Trayford20} to determine $\tau_\mathrm{ISM}$ and $\eta_\mathrm{ISM}$ from the dust surface density of each galaxy in \dustysage.

\cite{Lagos19} adopted the scaling of birth cloud optical depth with metal mass surface density from \cite{Lacey16} but modified it to use the dust surface density instead. However, since \textsc{SHARK} does not calculate dust properties of galaxies, they multiplied the gas-phase metal mass with an assumed dust-to-metal ratio to infer the total dust mass. In \mentari, we directly use the dust mass and gas mass to compute the optical depth,
\begin{equation}
    \tau_\mathrm{BC} = \tau_\mathrm{BC,0} \left[ \frac{f_\mathrm{DTG} \Sigma_\mathrm{gas,cl}}{f_\mathrm{DTM,MW} \Zsun \Sigma_\mathrm{MW,cl}} \right].
    \label{eq:tau_BC_lagos}
\end{equation}
Here, $\tau_\mathrm{BC,0} = 1$ is the default birth cloud optical depth in \cite{CF00}, $f_\mathrm{DTG} = M_\mathrm{dust} / M_\mathrm{gas}$ is the dust-to-gas mass ratio in \dustysage, and the cloud surface density $\Sigma_\mathrm{gas,cl}$ is defined as the maximum value between the cloud surface density of the Milky Way ($\Sigma_\mathrm{MW,cl} = 85 \Msun \mathrm{pc}^{-2}$) and the ISM gas surface density $\Sigma_\mathrm{gas}$. We use the dust half mass radius to compute the gas surface density in this equation. We adopt a solar metallicity of $\Zsun = 0.0189$ and the dust-to-metal ratio of the Milky Way of $f_\mathrm{DTM} = 0.33$. Using this formula, a typical spiral galaxy will have a birth cloud optical depth $\tau_\mathrm{BC} \approx \tau_\mathrm{BC,0}$. 

\subsubsection{Varying attenuation parameters following \citet{Somerville12}}
We have also tested the scaling relations proposed by \cite{Somerville12} to compute the optical depth of both the diffuse ISM and birth clouds. However, instead of using a metal mass surface density to determine the optical depth \citep[Equation 3 in][]{Somerville12}, we alter their prescription to use a dust mass surface density instead:
\begin{equation}
    \begin{split}
        \tau_\mathrm{ISM} &= \tau_0 \Sigma_\mathrm{dust} \\
        \tau_\mathrm{BC} &= \mu_\mathrm{BC} \tau_\mathrm{ISM}. 
    \end{split}
\end{equation}
$\tau_0$ and $\mu_\mathrm{BC}$ are treated as free parameters to match the observed ultraviolet luminosity function. We find a good agreement with $\tau_0 = 0.3$ and $\mu_\mathrm{BC} = 6.0$, not too far from the values used by \citet{Somerville12} ($\tau_0 = 0.2$ and $\mu_\mathrm{BC} = 4.9$).

\subsection{Infrared emission}
As described above, dust absorbs a fraction of the starlight in the ultraviolet to near-infrared. This energy increases dust temperature, which the dust grains reradiate in the mid to far-infrared. Therefore, the total infrared luminosity of a galaxy should be equal to the total absorbed starlight, known as the ``energy balance'' principle. The shape of the infrared SED correlates with the dust temperature, which is regulated by the grain size distribution and the intensity of the interstellar radiation field.

There are various techniques to construct an infrared SED template. Like the stellar spectral template, the infrared template can be constructed using synthetic or empirical spectra or some combination of both. Templates with compiled observed spectra often use only one or two parameters that drive the variation in the infrared spectra \citep[e.g.,][]{Rieke09, Dale14}. Meanwhile, templates with synthetic spectra have more free parameters that can change the infrared spectral shape, usually related to the physical properties of the dust and interstellar radiation field \citep[e.g.,][]{DL07, daCunha08}. This work explores how different approaches in computing infrared SEDs result in the spectral shape and galaxy luminosity function.

We provide two infrared templates developed using different approaches in \mentari. The first is the two-dimensional template of \cite{Dale14} which is built on a semi-empirical model. In the template, a local SED is constructed using three dust components: large grains, ``very small grains", and polycyclic aromatic hydrocarbons (PAHs). The thermal properties of each component depends on their size distribution. Large grains are assumed to be in thermal equilibrium, so their emission is modelled using a modified black body spectrum, contributing largely to the far-infrared regime. ``Very small grains'' are stochastically heated and are responsible for the broad mid-infrared continuum. PAHs are molecules excited by a single UV photon on their CH vibrational, stretching and bending modes. Hence, they emit in specific broad lines associated with those modes. The \cite{Dale14} model assumes two compositions for grains: graphite and silicate. 

The underlying physics of the model is that a mass of dust ($dM_\mathrm{d}$) is heated in `local' radiation fields with an intensity range of $0.3 \le U \le 10^5$, with $U=1$ representing the heating intensity of the Solar Neighborhood. The total infrared SED is the average of all local SEDs comprising it, combined using a power-law equation:
\begin{equation}
    dM_\mathrm{d} \propto U^{-\alpha_\mathrm{SF}} dU.
\end{equation}
$\alpha_\mathrm{SF}$ is the parameter that determines the contribution of the various local SEDs. This model is calibrated using a set of observational constraints. The \cite{Dale14} template also parametrises the fraction of AGN emission, which we currently fix at zero. 

When deciding which spectral set to be applied in a galaxy, \mentari adopts the correlation between the $\alpha_\mathrm{SF}$ parameter with the total infrared luminosity ($L_\mathrm{IR}$) of the galaxy. From the energy balance principle, the total infrared luminosity is computed from the total attenuated luminosity:
\begin{equation}
    L_\mathrm{IR} = \int_{912\text{\AA}}^{\infty} (L_\lambda^0 - L_\lambda) d\lambda,
    \label{eq:LIR}
\end{equation}
where $L_\lambda^0$ is the intrinsic stellar spectrum before attenuation and $L_\lambda$ is the spectrum after applying an attenuation curve. We use the correlation provided in \cite{Rieke09} to compute $\alpha_\mathrm{SF}$ from $L_\mathrm{IR}$:
\begin{equation}
    \alpha_\mathrm{SF} = 10.096 - 0.741 \log L_\mathrm{IR}.
\end{equation}
Because the \citet{Rieke09} relation is capped at $\log L_\mathrm{IR} = 11.6$, we assume that galaxies with total infrared luminosity higher than this cap to have the same $\alpha_\mathrm{SF}$ with those with $\log L_\mathrm{IR} = 11.6$. 

The second template used in \mentari is presented in \cite{Safarzadeh16}. This template comprises far-infrared SEDs built using the 3D dust radiative transfer code \textsc{SUNRISE} \citep{Jonsson10} and a hydrodynamical simulation suite \citep{Hayward11, Hayward12, Hayward13, Hayward14, Lanz14}. The coupling of a radiative transfer code with the output of a hydrodynamical simulation currently provides the most rigorous description of the infrared radiation process. A set of radiative transfer equations considers various grain properties and can accurately model the geometry between stars and dust, dust scattering and absorption, and dust re-emission. However, running such extensive calculations requires substantial computational power. Therefore, \cite{Safarzadeh16} made an effort to parameterise the theoretical SEDs. 

\cite{Safarzadeh16} grouped the simulated SEDs based on their infrared luminosity and dust mass, resulting in a total of 22 grids. Then, they computed the median SED for each grid (see their Figure 11). The infrared luminosity and dust mass are chosen as the main parameters based on an evaluation using a principal component analysis (PCA) technique. They explored the influence of several galaxy properties on the shape of far-infrared SEDs. The properties include star formation rate, total infrared luminosity, AGN luminosity, and dust mass. The total infrared luminosity and dust mass are chosen as the model parameters because the authors found that they are the two most important factors determining the infrared spectral shape.

\begin{figure*}
    \centering
    \includegraphics[width = 1.0\textwidth]{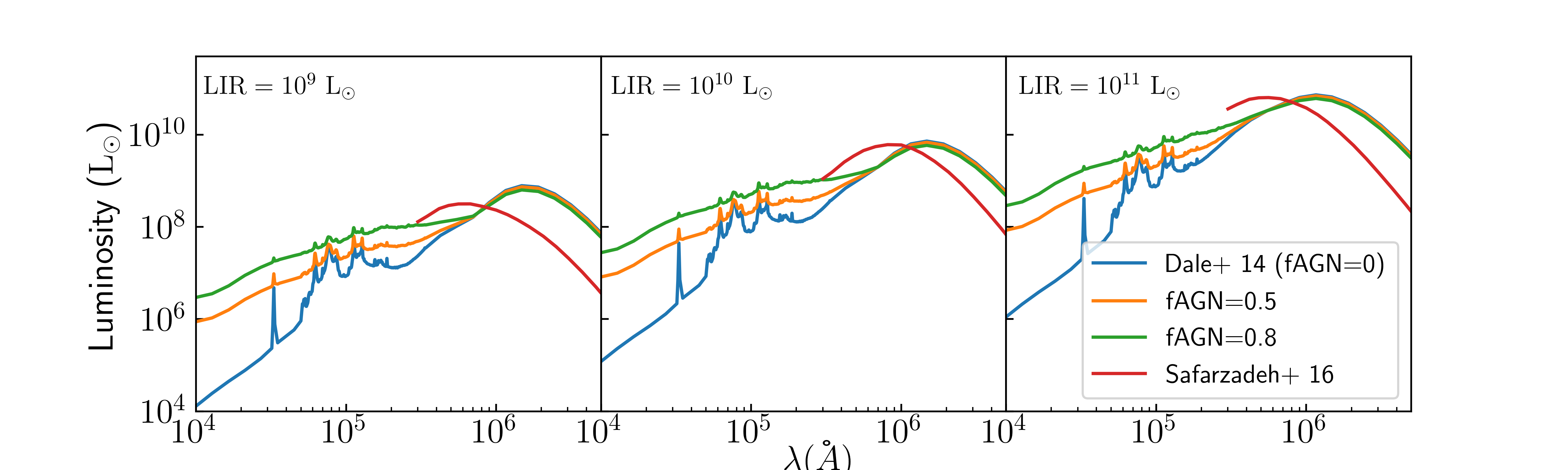}
    \caption{Comparison of the dust emission templates of \citet{Dale14} and \citet{Safarzadeh16}. The blue, orange, and green lines are from \citet{Dale14} templates with AGN fraction $f_\mathrm{AGN} = 0, 0.5$, and $0.8$ as indicated in the legend. The red line is from \citet{Safarzadeh16} template. Note that this template only provide spectra in the far-infrared regime. The three panels show the templates for total infrared luminosities of $10^{9}$ \Lsun, $10^{10}$ \Lsun, and $10^{11}$ \Lsun.}
    \label{fig:IR_templates}
\end{figure*}
 
\mentari uses the total infrared luminosity computed using Equation \ref{eq:LIR} and the computed ISM dust mass from \dustysage to decide which grid to use from the \cite{Safarzadeh16} template. Figure \ref{fig:IR_templates} shows a comparison of infrared SEDs from the \cite{Dale14} template and that of \cite{Safarzadeh16}. In the mid-infrared, we only plot the \cite{Dale14} spectra since the \cite{Safarzadeh16} template only provides far-infrared spectra. The template from \citet{Dale14} takes the AGN fraction ($f_\mathrm{AGN}$) and total infrared luminosity (LIR) as free parameters. $f_\mathrm{AGN}$ determines the mid-infrared part of the spectrum while LIR affects the normalisation. In this work, we use $f_\mathrm{AGN}=0$ for all model variants so our mid-infrared results should be taken as a lower limit. The \citet{Safarzadeh16} template has the total LIR and dust mass as its parameters. At a fixed LIR, the luminosity in both templates reflects very different grain temperature distributions. 

We chose these two templates because they only have two parameters, which minimise the uncertainty that arises from having to fix multiple parameters that are not directly available from \dustysage and \mentari. \citet{Dale14} has been used extensively in previous SED modelling codes and the IR SED is mainly driven by the total infrared luminosity, which we can compute using the energy balance principle in \mentari. We pick the \citet{Safarzadeh16} template because it is derived theoretically, offering a different approach than \citet{Dale14}. It also has only two parameters: the total infrared luminosity, which we derive from \mentari, and the total dust mass which we take from the output of \dustysage.

\subsection{Model variants}
\label{ssec:variants}

We explore a variety of prescriptions and templates with \mentari to test our ability to reproduce several key observations. Our model variants are described in Table \ref{tab:variant}. We change several parameters in each variant, focusing on the dust attenuation and infrared emission prescriptions. We provide a model variant with pure stellar spectra and no attenuation nor infrared template, called the ``Unattenuated'' variant. We adopt three attenuation formulas in other model variants. Our ``Default'' model uses the scaling for the \citet{CF00} dust attenuation parameter from \citet{Lagos19} (which follows the birth cloud dust scaling from \citet{Lacey16} and the diffuse ISM dust scaling from \citet{Trayford20}). 

We also explore the scaling for birth clouds and the ISM optical depth from \citet{Somerville12} in the ``Somerville'' model variant. In this variant, we apply $\eta_\mathrm{BC} = -1.3$ proposed by previous authors \citep{daCunha08, Wild11} to match the observed ultraviolet luminosity function better. For our ``CF00'' variant, we adopt the original parameter set of \citet{CF00}. When alternating attenuation parameters, we always utilise the mid-infrared template of \citet{Dale14} and the far-infrared template of \citet{Safarzadeh16} to keep consistency. In all models, we use $\eta_\mathrm{ISM} = -0.7$ from \citet{CF00}.

In addition to exploring several attenuation prescriptions, we also investigate two far-infrared templates: \citet{Safarzadeh16} (the ``Default'' model) and \citet{Dale14} (the ``Dale'' model). When alternating infrared templates, we use the attenuation scaling from \citet{Lagos19} and the mid-infrared templates from \citet{Dale14}. Table \ref{tab:variant} describes each model variant and specifies the line style for each variant in the following figures.

\begin{table*}
    \caption{Summary of the SED model parameters explored in this work.}
    \centering
    \resizebox{1.0\textwidth}{!}{%
    \begin{tabular}{lccccccc}
        \hline
        \hline
        name & $\tau_\mathrm{BC}$ & $\eta_\mathrm{BC}$ & $\tau_\mathrm{ISM}$ & $\eta_\mathrm{ISM}$ & far-IR template & line style \\
        \hline
        \textbf{Varying attenuation} &&&&&& \\
        Unattenuated & - & - & - & - & - & cyan solid \\
        Default  & \citet{Lagos19} & -0.7 & \citet{Trayford20} & \citet{Trayford20} & \citet{Safarzadeh16} & black solid \\ 
        Somerville & \citet{Somerville12} & -1.3 & \citet{Somerville12} & -0.7  & \citet{Safarzadeh16} & dotted red \\ 
        CF00 & 1.0 & -0.7 & 0.3 & -0.7 & \citet{Safarzadeh16} & dash-dotted blue \\
        \hline
        \textbf{Varying infrared} &&&&&& \\
        Default  & \citet{Lagos19} & -0.7 & \citet{Trayford20} & \citet{Trayford20} & \citet{Safarzadeh16} & black solid \\
        Dale & \citet{Lagos19} & -0.7 & \citet{Trayford20} & \citet{Trayford20} & \citet{Dale14} & dashed green \\
        \hline
        \hline
    \end{tabular}
    }
    \label{tab:variant}
\end{table*}

\section{Galaxy emission from far-ultraviolet to far-infrared}
\label{sec:results}
In this section, we present galaxy emission from the \dustysage semi-analytic model combined with the \mentari SED generator. We want to test how our various model variants compare with the observations of the global galaxy population. First, in Section \ref{ssec:z0LF} we investigate the $z=0$ luminosity functions from the far-ultraviolet to far-infrared. Then, in Section \ref{ssec:csed} we present the cosmic SED (CSED) at redshift $z=0$. In Section \ref{ssec:zLF} we predict the evolution of the luminosity function from redshift $z=0.5$ to $z=3$. All luminosity functions are shown in the rest-frame wavelength at the specified redshift. 

\subsection{Dust contribution on panchromatic emission at \mbox{$z=0$}}
\label{ssec:z0LF}

Emission of a galaxy in different wavelength provide us information about the properties of its constituent. In this work, we are testing how dust affect the galaxy emission across the panchromatic window. To achieve this, we compare the luminosity functions and CSED of model galaxies constructed using different dust prescriptions listed in Table \ref{tab:variant}. We also discuss the contribution of other galaxy components such as stars and AGN when appropriate.

\begin{figure}
    \centering
    \includegraphics[width=1.1\linewidth]{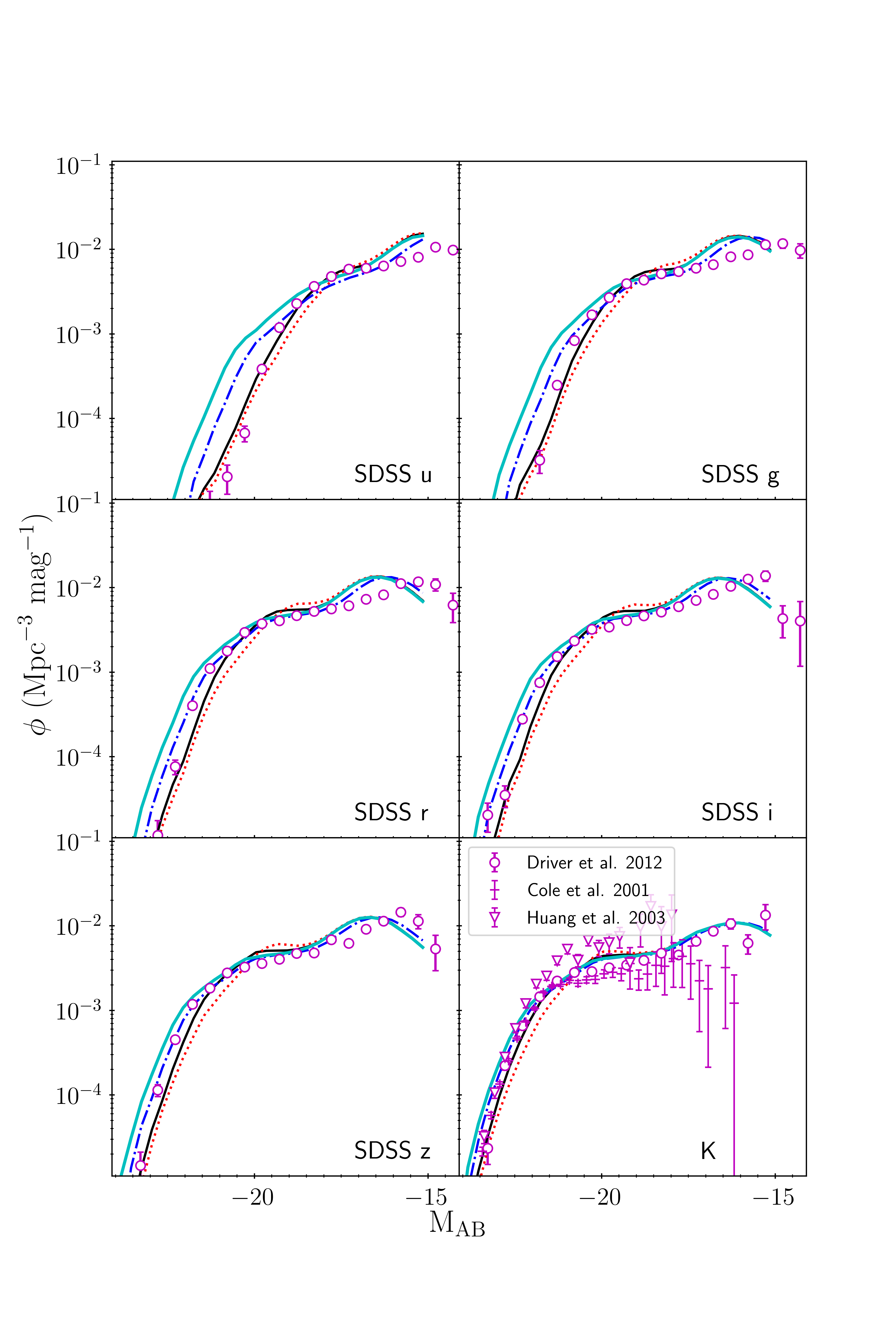}
    \caption{Luminosity functions at $z=0$ for the SDSS \textit{ugriz} and K-bands. We show four model variants as defined in the upper part of Table \ref{tab:variant}: the cyan, black, red dotted and blue dash dotted lines are our Unattenuated, Default, Somerville and CF00 models, respectively. The symbols with error bars are the observational values from \citet{Driver12}, \citet{Cole01} and \citet{Huang03}, as indicated in the legend.}
    \label{fig:optical}
\end{figure}

We present our predicted SDSS and K-band luminosity functions at $z=0$ in Figure \ref{fig:optical}. These bands are mainly dominated by direct stellar radiation and where dust attenuation is simplest. The observational values for the SDSS bands are taken from \citet{Driver12}. In these bands, all models provide excellent agreement with the observations, which is expected because \dustysage is tuned to match the $z=0$ stellar mass function. The gap between models with different attenuation parameters decreases towards the K-band, demonstrating the decrease of attenuation with increasing wavelength. In the last panel, we compare our prediction for the $z=0$ K-band luminosity function with \citet{Driver12, Cole01} and \citet{Huang03}. Regardless of the attenuation parameters, all model variants produce similar results, which shows that dust attenuation has a negligible effect on galaxy emission here. 

\begin{figure}
    \centering
    \includegraphics[width=1.1\linewidth]{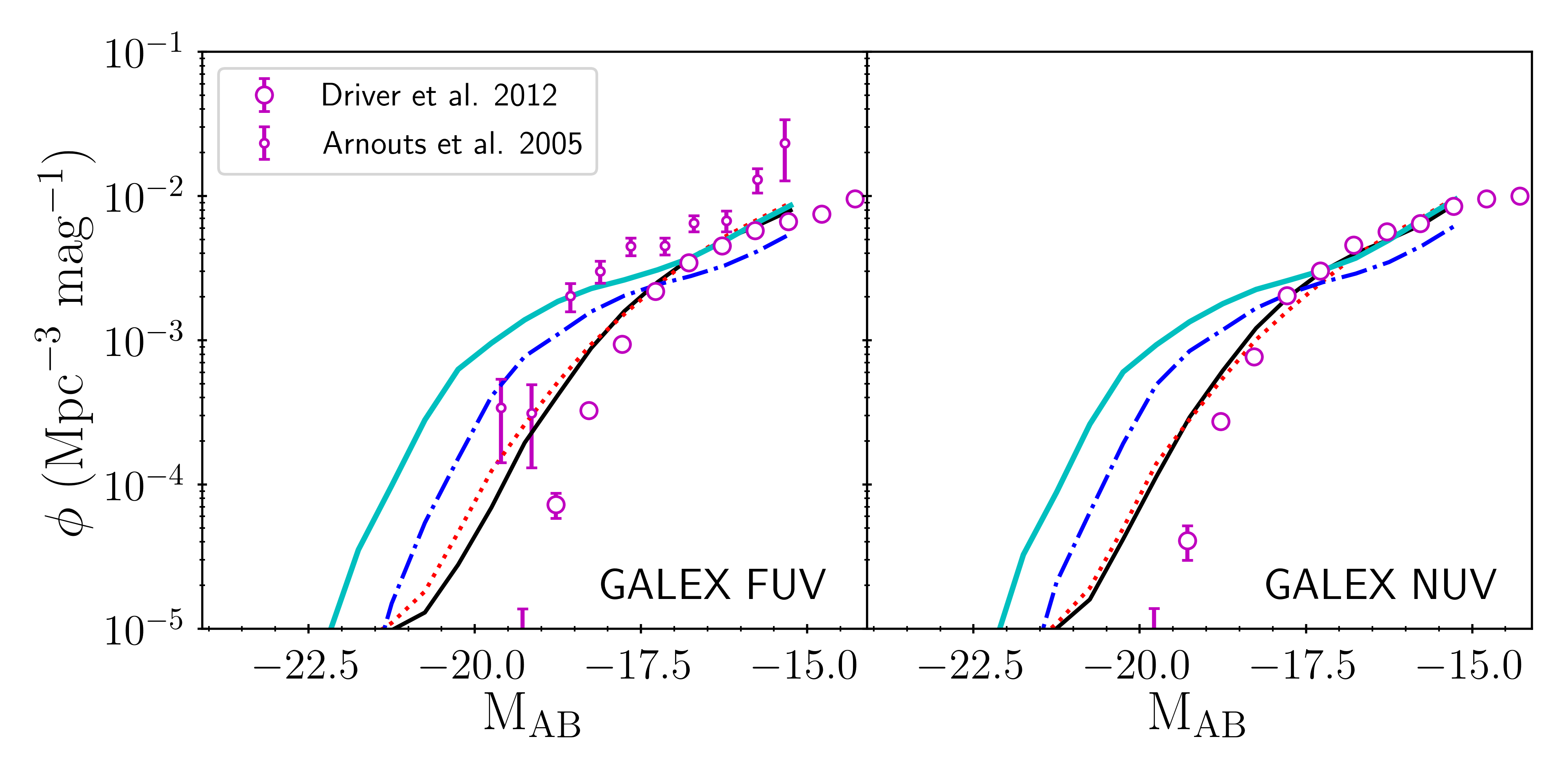}
    \caption{Luminosity functions at $z=0$ for the GALEX FUV and NUV bands. We show four model variants as defined in Table \ref{tab:variant}: the cyan, black, red dotted and blue dash dotted lines are our Unattenuated, Default, Somerville and CF00 models, respectively. The symbols with error bars are the observational values from \citet{Driver12} and \citet{Arnouts05}, as indicated in the legend.}
    \label{fig:UV}
\end{figure}

Figure \ref{fig:UV} shows the comparison of our $z=0$ luminosity functions for the GALEX far-ultraviolet (FUV) and near-ultraviolet (NUV) bands with the observational results of \citet{Driver12} and \citet{Arnouts05}. The luminosity function at these bands is primarily a test of the amount of attenuated star formation in the $z=0$ Universe. We present the prediction from our four model variants with different attenuation prescriptions described in Table \ref{tab:variant}. As expected, the Unattenuated model (solid cyan lines) gives the worst agreement with the observations. Compared to the observational datasets from \citet{Arnouts05} and \citet{Driver12} at the far and near-ultraviolet, the model with CF00 constant attenuation parameters (dash-dotted blue lines) overproduces bright galaxies with AB magnitudes of $M_\mathrm{AB} \ge -18$, indicating that the attenuation in this variant is too small. In the far-ultraviolet, scaling the attenuation parameters with the computed dust masses from \dustysage, represented by our Default model (solid black lines) and Somerville model (dotted red lines), improves the match with the datasets from \citet{Arnouts05} and \citet{Driver12}. Although these models still overproduce the number density of the brightest galaxies compared to \citet{Arnouts05}. This overproduction occurs in the near-ultraviolet waveband as well. 

We now move to the wavelength range where stellar emission starts to become insignificant. The mid-infrared is contributed by both dust emission and AGN. We investigate how both constituents affect the luminosity function at different filters. 

\begin{figure}
    \centering
    \includegraphics[width=1.1\linewidth]{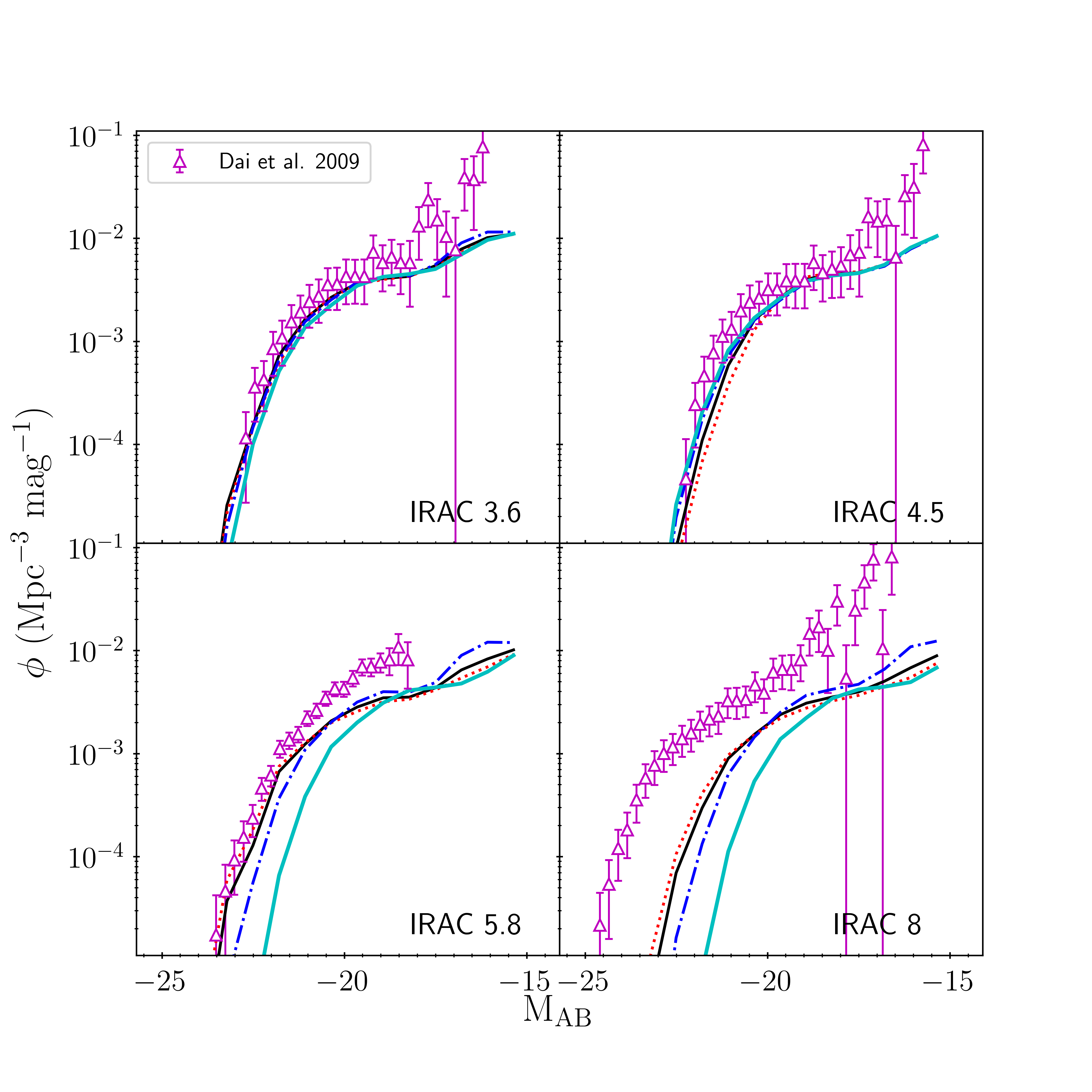}
    \caption{Luminosity functions at $z=0$ for the IRAC $3.6 \mu \mathrm{m}$, $4.5 \mu \mathrm{m}$, $5.8 \mu \mathrm{m}$ and $8.0 \mu \mathrm{m}$ bands. We show four model variants with different attenuation formulas as described in Table \ref{tab:variant}: the cyan, black, red dotted and blue dash dotted lines are our Unattenuated, Default, Somerville and CF00 models, respectively. The symbols with error bars are the observational values from \citet{Dai09}, as indicated in the legend.}
    \label{fig:mir_a}
\end{figure}

Figure \ref{fig:mir_a} presents our predicted $z=0$ mid-infrared luminosity functions for the Spitzer IRAC $3.6 \mu \mathrm{m}$, $4.5 \mu \mathrm{m}$, $5.8 \mu \mathrm{m}$ and $8.0 \mu \mathrm{m}$ bands. Here, we compare with the measurement of \citet{Dai09}. Figure \ref{fig:mir_a} shows the variations driven by the attenuation models, which are not as significant as in the ultraviolet. Here, we use the same mid-infrared templates from \citet{Dale14} on all model variants. In the $3.6 \mu \mathrm{m}$ and $4.5 \mu \mathrm{m}$ restframes, our models provide a good match with the observations. However, all of our model variants underproduce the galaxy emission in the IRAC $5.8 \mu \mathrm{m}$ and $8.0 \mu \mathrm{m}$ bands. 

The mismatch in the mid-infrared is particularly complicated. Figure \ref{fig:IR_templates} shows that the mid-infrared SEDs contain aromatic features from the PAH molecules and are influenced heavily by the AGN fraction parameter. The mid-infrared emission from AGN comes from reprocessing the energetic photons by its dusty torus \citep{Franceschini02, Dwek13}; hence, increasing the AGN fraction will enhance the mid-infrared flux. \citet{Leja18} reveal that at least $10 \%$ of the mid-infrared flux in local galaxies is contributed by galaxies that host AGN. The accounting of AGN emission is particularly crucial because a sizable proportion of the galaxy population has significant AGN components. $35 \%$ of SDSS galaxies are classified to contain composite or AGN source based on the Baldwin-Phillips-Terlevich (BPT) diagram \citep{Kauffmann03}. Using SED-fitting, \citet{Thorne22} detected significant AGN emission ($f_\mathrm{AGN} > 0.1$) in $28.7 \%$ of galaxies in GAMA survey and $41.5 \%$ in the DEVILS survey. Although, the percentage in DEVILS should be treated as an upper limit due to its incomplete far-infrared data. 

Our model (Figure \ref{fig:mir_a}) adopts $f_\mathrm{AGN} = 0$ when applying the infrared template from \citet{Dale14}, so it should be treated as the lower limit in the IRAC $5.8 \mu \mathrm{m}$ and $8.0 \mu \mathrm{m}$ bands where AGN contamination is non-negligible. In addition, the scaling relation from \citet{Rieke09} that we use to convert the total LIR from \mentari to the $\alpha$-parameter in the \citet{Dale14} template is based on the observation of pure star-forming galaxies only. A future improvement of this work will include using a realistic AGN fraction and modified scaling relation appropriate for AGN host galaxies to generate infrared spectra.

\begin{figure}
    \centering
    \includegraphics[width=1.1\linewidth]{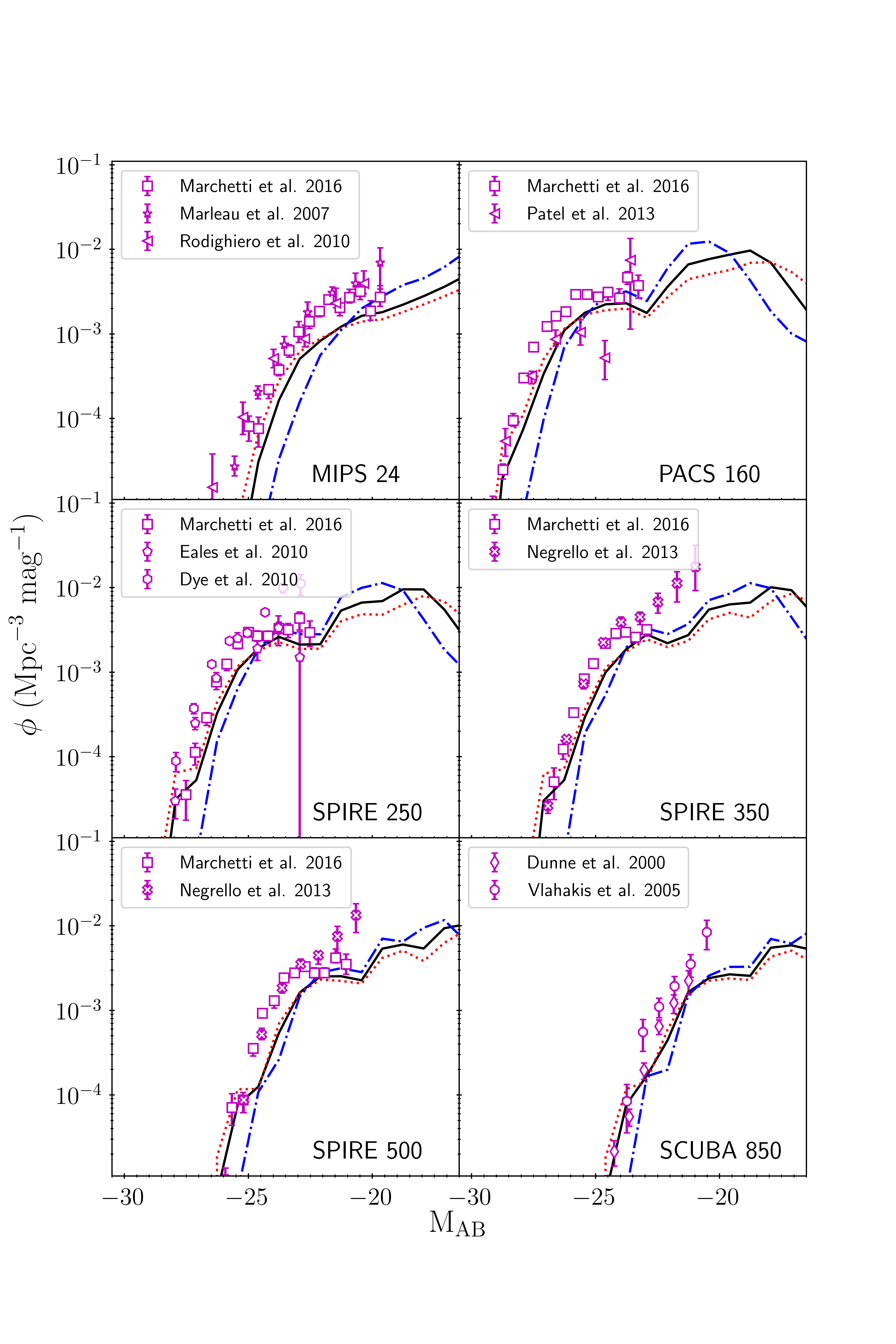}
    \caption{Luminosity functions at $z=0$ for the MIPS $24 \mu \mathrm{m}$, PACS $160 \mu \mathrm{m}$, SPIRE $250 \mu \mathrm{m}$, $350 \mu \mathrm{m}$, $500 \mu \mathrm{m}$ and SCUBA $850 \mu \mathrm{m}$ bands. We show three model variants with different attenuation formulas as described in Table \ref{tab:variant}: the black, red dotted and blue dash dotted lines are our Default, Somerville and CF00 models, respectively. The unattenuated model doesn't extend to this wavelength regime. The symbols with error bars are the observational values from \citet{Marchetti16}, \citet{Marleau07}, \citet{Rodighiero10}, \citet{Patel13}, \citet{Eales10}, \citet{Dye10}, \citet{Negrello13}, \citet{Dunne00} and \citet{VDE05}, as indicated in the legend.}
    \label{fig:fir_a}
\end{figure}

\begin{figure}
    \centering
    \includegraphics[width=0.9\linewidth]{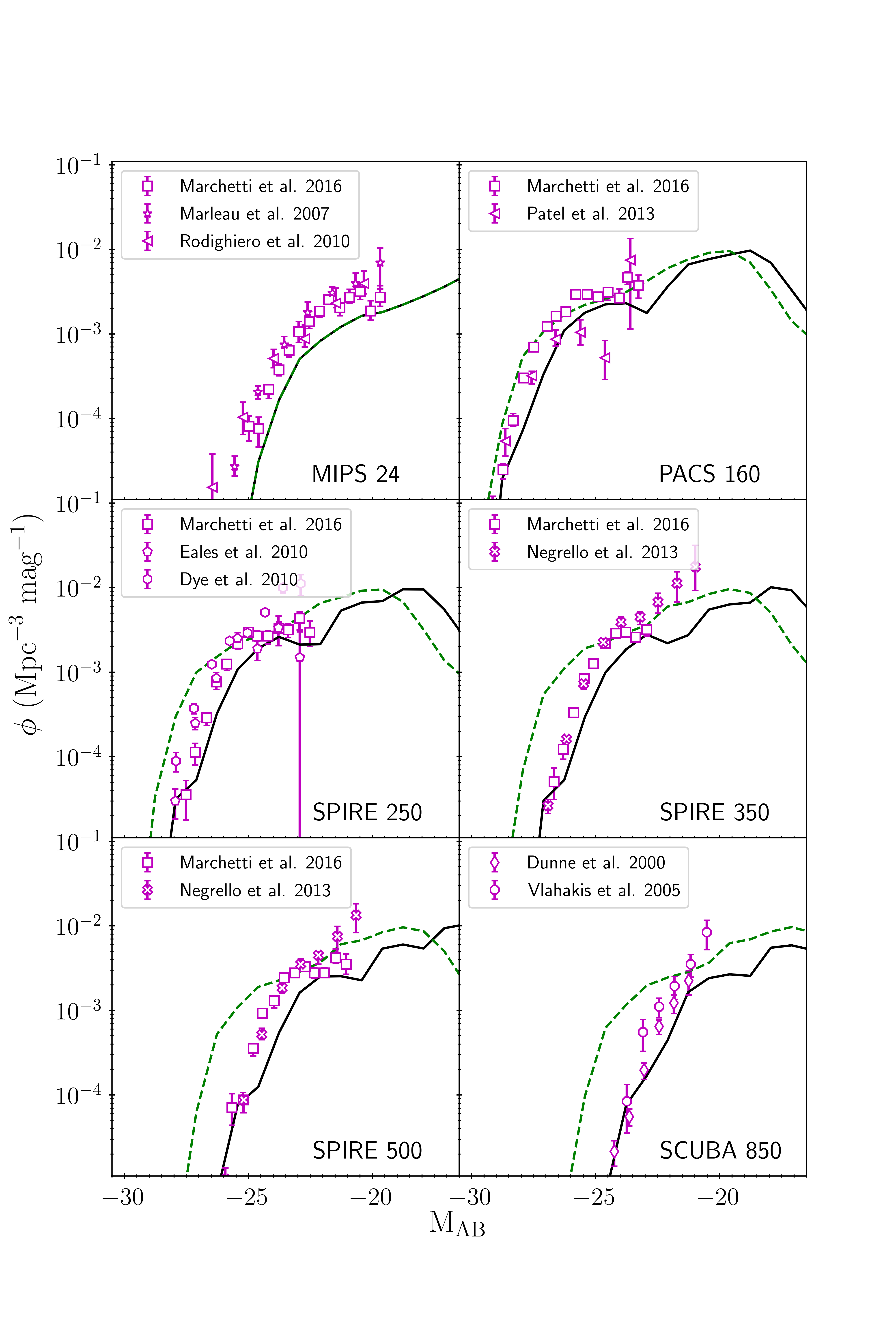}
    \caption{Luminosity functions at $z=0$ for the MIPS $24 \mu \mathrm{m}$, PACS $160 \mu \mathrm{m}$, SPIRE $250 \mu \mathrm{m}$, $350 \mu \mathrm{m}$, $500 \mu \mathrm{m}$ and SCUBA $850 \mu \mathrm{m}$ bands. We show two model variants with different infrared templates as described in Table \ref{tab:variant}: the black lines represent our Default model and the green dashed lines mark the Dale model. The symbols with error bars are the observational values from \citet{Marchetti16}, \citet{Marleau07}, \citet{Rodighiero10}, \citet{Patel13}, \citet{Eales10}, \citet{Dye10}, \citet{Negrello13}, \citet{Dunne00} and \citet{VDE05}, as indicated in the legend.}
    \label{fig:fir_b}
\end{figure}

We present the $z=0$ far-infrared luminosity functions in Figure \ref{fig:fir_a} and \ref{fig:fir_b}, including the MIPS $24 \mu \mathrm{m}$; PACS $160 \mu \mathrm{m}$; SPIRE $250 \mu \mathrm{m}$, $350 \mu \mathrm{m}$, $500 \mu \mathrm{m}$; and SCUBA $850 \mu \mathrm{m}$ bands. Observational values are taken from \citet{Marchetti16, Marleau07, Rodighiero10, Patel13, Eales10, Dye10, Negrello13, Dunne00}, and \citet{VDE05}. In these bands, dust re-emission of the absorbed light in the ultraviolet to optical bands dominate the spectrum. Figure \ref{fig:fir_a} shows how the variation in far-infrared bands are driven by the attenuation prescriptions while Figure \ref{fig:fir_b} shows the variation is driven by the different far-infrared templates. Note that we do not plot the unattenuated emission here because the pure stellar spectra does not extend to these wavelength ranges. 

Figure \ref{fig:fir_a} shows that across the far-infrared restframe, our Default and Somerville model variants (solid black and dotted red lines, respectively) provide a good fit within the error bars of the observed datasets. These variants use the dust mass from \dustysage to compute the attenuation parameters of birth clouds and diffuse ISM. On the other hand, the CF00 model (dash-dotted blue line) that uses constant attenuation parameters adopted from \citet{CF00} underpredicts the number of bright galaxies across these bands. The ultraviolet and optical luminosity functions (Figure \ref{fig:UV} and \ref{fig:optical}) show that the CF00 model gives the smallest amount of attenuation compared to the other models, which explains the lack of emission in the infrared wavebands. This supports our conclusion for the need to compute attenuation parameters self-consistently to reproduce the observed far-infrared emission from galaxies.

In Figure \ref{fig:fir_b}, we plot the Default model that uses the \citet{Safarzadeh16} far-infrared template (black solid lline) and the Dale model (green dashed line). While in MIPS $24 \mu \mathrm{m}$ both models slightly underpredicts the observational data, in all other panels the data lies between the Default model and the Dale model. In general, the Dale model is in better agreement with the observation below the knee of the luminosity functions, where the Default model slightly underproduces the emission. However, the Dale model overpredicts the number density for the bright infrared galaxies above the knee, and the observational points lie closer to the Default model. 

Overall, we have presented how our model variants predict the $z=0$ luminosity function ranging from the far-ultraviolet to far-infrared. The CF00 model variant with constant attenuation parameters slightly overestimates the ultraviolet emission and systematically underpredicts the number of bright far-infrared galaxies while reproducing the optical to near-infrared galaxy emission reasonably well. This tension in the far-infrared is found with many previous SAMs \citep{Baugh05, Lacey16, Somerville12}. Some suggest that varying the IMF is a necessary solution. We find that adopting a dust mass computed rigorously in a SAM to calculate the attenuation parameters solves this problem without the need to invoke a varying IMF. The same result is shown in \citet{Lagos19}. These authors use the observed dust-to-metal ratio from \citet{RR14} to compute dust mass for each galaxy from its metal mass, then derive the attenuation based on the inferred dust mass.

\subsection{The local CSED}
\label{ssec:csed}

\begin{figure*}
    \centering
    \includegraphics[width=0.8\textwidth]{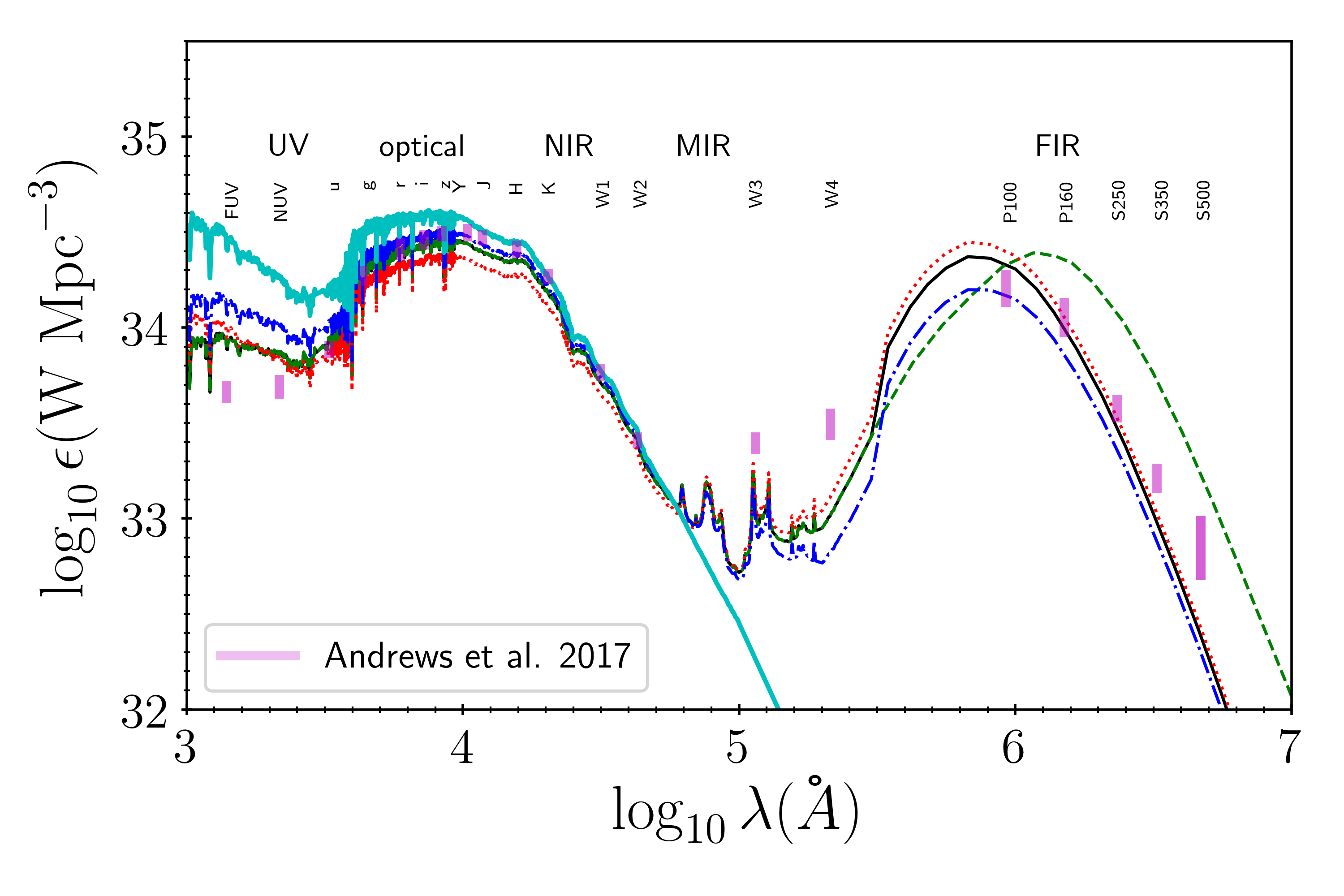}
    \caption{Our predicted Cosmic Spectral Energy Distribution at $z=0.0$. We show four model variants as defined in Table \ref{tab:variant}: the cyan, black, red dotted and blue dash dotted lines are our Unattenuated, Default, Somerville and CF00 models, respectively. The green dashed lines mark the Dale model which use a different far-infrared template to the Default model (black line). The shaded areas are the observational values from \citet{Andrews17}, as indicated in the legend.}
    \label{fig:csed}
\end{figure*}

This section discusses galaxies' cosmic spectral energy distribution (CSED) at $z=0$. To compute the CSED, we add the restframe spectra of every single galaxy from \dustysage at $z=0$ and normalise based on the co-moving volume of the simulation. In Figure \ref{fig:csed}, we compare the local CSED from our four model variants (see Table \ref{tab:variant}) with the observed CSED of \citet{Andrews17} using the GAMA survey \citep{Driver09}.

In general, our results show a good agreement with the observed CSED across the ultraviolet to far-infrared wavelengths, but each variant performs differently. Our Somerville model gives the best match in the far and near-ultraviolet, while our Default model overestimates the emission by $0.05 - 0.1$ dex. The overestimation is more significant for the CF00 model, which is expected because, as we have shown, the constant attenuation parameters used in this variant provide less attenuation than the other models. The excellent match of the Somerville model variant is also expected because we scale the attenuation parameters here to match the ultraviolet luminosity function. 

Across the SDSS u and g band wavelengths, all model variants agree well with the observations. The differences between the model variants are less than that seen in the far and near-ultraviolet bands but still show the critical effects of attenuation. Our model provides excellent agreement with the observed CSED at optical wavelengths. However, we systematically underpredict the emission closer to the near-infrared, although the gap is only $\sim 0.05$ dex. This systematic difference is also found in galaxies from the \textsc{EAGLE} simulation \citep{Baes19}. 

In the mid-infrared, our model performs poorly. As mentioned above, the mid-infrared emission is particularly difficult due to the many complicated ingredients of the spectrum, including PAH (see Figure \ref{fig:IR_templates}) and AGN emission \citep{Franceschini02, Dwek13}. Because we lock the AGN fraction in our model at zero, our prediction should be treated as lower limit. In the far-infrared region, especially below $250 \mu \mathrm{m}$, our Default model prediction lies within the uncertainties of the observed measurements. In this regard, our model variants using the \textsc{SUNRISE} infrared templates \citep{Safarzadeh16} better match the observations compared to that with the infrared templates of \citet{Dale14} (green dashed line), which overestimate the observations by $\sim 0.35$ dex. This could be caused by the different parameters used in both templates. The main parameters for the \citet{Safarzadeh16} templates are the total infrared luminosity and dust mass. We have computed dust mass directly from \dustysage, and we assume the total infrared luminosity directly from the total attenuated spectra. On the other hand, \citet{Dale14} use the $\alpha$ parameter and AGN fraction as their main parameters. We fix the AGN fraction to zero when incorporating the template into \mentari. Then, we use the relation presented in \citet{Rieke09} to compute the $\alpha$ parameter from the total infrared luminosity, which might add uncertainty to the results. However, at the longest wavelengths, $\ge 500 \mu \mathrm{m}$, our \citet{Safarzadeh16} model variants underestimate the observed values by $\sim 0.3$ dex. \citet{Baes19} also found the same tension with their EAGLE data. This is possibly due to the lack of observational measurements beyond $24 \mu \mathrm{m}$, so the compiled CSED of \citet{Andrews17} could be poorly constrained.

Overall, Figure \ref{fig:csed} shows that our model has succeeded in extending the semi-analytic model prediction for the galaxy SED to both the ultraviolet and infrared ends. This is a major advance considering our predecessor, \sage, and many previous SAMs only provide predictions in the optical to near-infrared bands, from SDSS u to the K-band, where the stellar emission dominates. Our rigorous dust treatment has allowed us to achieve a more realistic prediction naturally, without the need to invoke more exotic solutions such as a varying IMF, as suggested by previous works. 

\subsection{Redshift evolution of the luminosity functions}
\label{ssec:zLF}

The $z=0$ luminosity functions test the ability of \mentari to derive the observed SED from the fundamental properties of galaxies (e.g., stellar mass, age, metallicity, dust mass). However, we are also interested in understanding the critical processes across galaxy history that result in their present properties. Comparing our predicted redshift evolution with the observations will reflect how successful the model is in implementing the complicated physical processes of galaxy formation. However, please note that the \dustysage SAM used in this work is only calibrated using observational constraints from local galaxies. Therefore, our predictions at high redshift are not meant to reproduce the observational value but rather to serve as a tool to understand the complex behaviour of stars and dust in producing the galaxy emission across cosmic time. 

\subsubsection{Effects of attenuation prescription}
\label{sssec:attenuation-effect}

\begin{figure}
    \centering
    \includegraphics[width=1.1\linewidth]{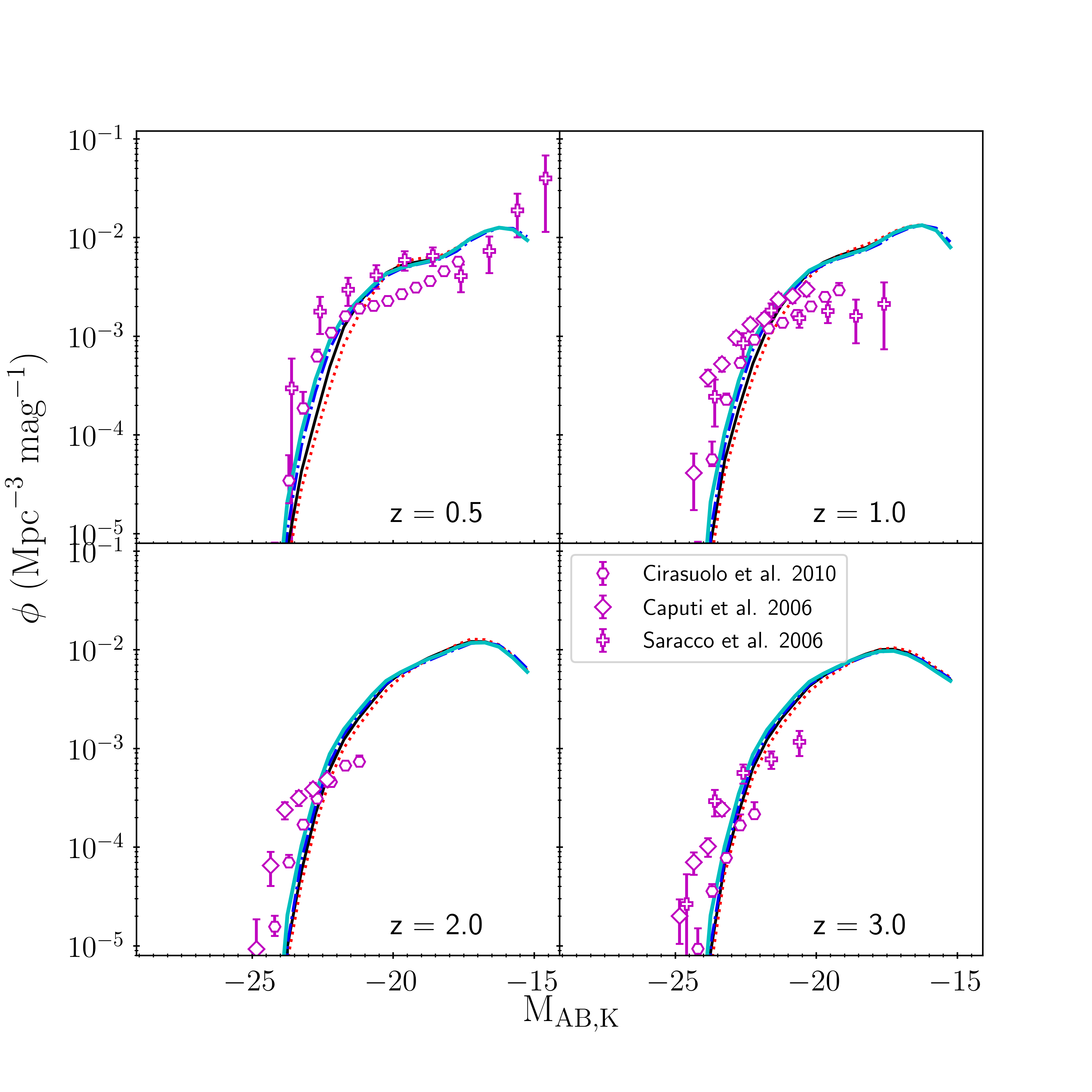}
    \caption{Our predicted K-band luminosity functions at $z=0.5, 1.0, 2.0, 3.0$. We show four model variants with different attenuation formula as described in Table \ref{tab:variant}: the cyan, black, red dotted and blue dash dotted lines are our Unattenuated, Default, Somerville and CF00 models, respectively. The symbols with error bars are the observational values from \citet{Cirasuolo10}, \citet{Saracco06}, and \citet{Caputi06}, as indicated in the legend.}
    \label{fig:K-evol}
\end{figure}

In this section, we analyse our results based on the model variants described in section \ref{ssec:variants}. In Figure \ref{fig:K-evol}, we show the evolution of the K-band luminosity function out to $z=3$. We can see that all model variants provide similar values, reflecting that the attenuation effect is not essential in this band. We compare our SAM with the observed measurements of \citet{Cirasuolo10, Saracco06} and \citet{Caputi06}. At $z=0.5$, we find excellent agreement with the observed values. From $z=1$ to $z=3$, our predictions are not as good as those at the lower redshifts but agree reasonably well with the observations. This is surprising given the free parameters in \dustysage were only chosen to match the galaxy properties at $z=0$. 

At $z\ge1$, we slightly overproduce the number of galaxies below the knee of the K-band luminosity function. Our results are similar to the findings of many previous SAMs \citep{Fontanot09, Cirasuolo10, Henriques11, Somerville12}. A study by \citet{Fontanot09} explored three independent SAMs and found that all produced lower mass galaxies too early, resulting in an overabundance of faint galaxies at high redshift. Introducing a scaling for the reincorporation timescale of ejected gas with redshift and galaxy/halo properties has been found to resolve this tension in \citet{Henriques13}. However this ``solution'' does not work in \sage and \dustysage, or similarly the Somerville SAM (Rachel Somerville, private communication). So unfortunately appears unique to the Henriques model only and the way it was constructed.

At $z=2$ and $z=3$, our prediction for the brightest galaxies ($M_\mathrm{AB,K} \le -24$) are somewhat lower than the observed values. The low contribution of Asymptotic Giant Branch (AGB) stars in the \citet{2003BC} stellar population model might be responsible for this difference. The AGB phase of stellar evolution is poorly mapped, but it significantly affects the near-infrared flux. Previous studies have found that using the \citet{Maraston05} stellar population model in a SAM can resolve the number discrepancy of the brightest K-band galaxies at $z=2$ and $z=3$ \citep{Henriques10, GP14}.

\begin{figure}
    \centering
    \includegraphics[width=1.1\linewidth]{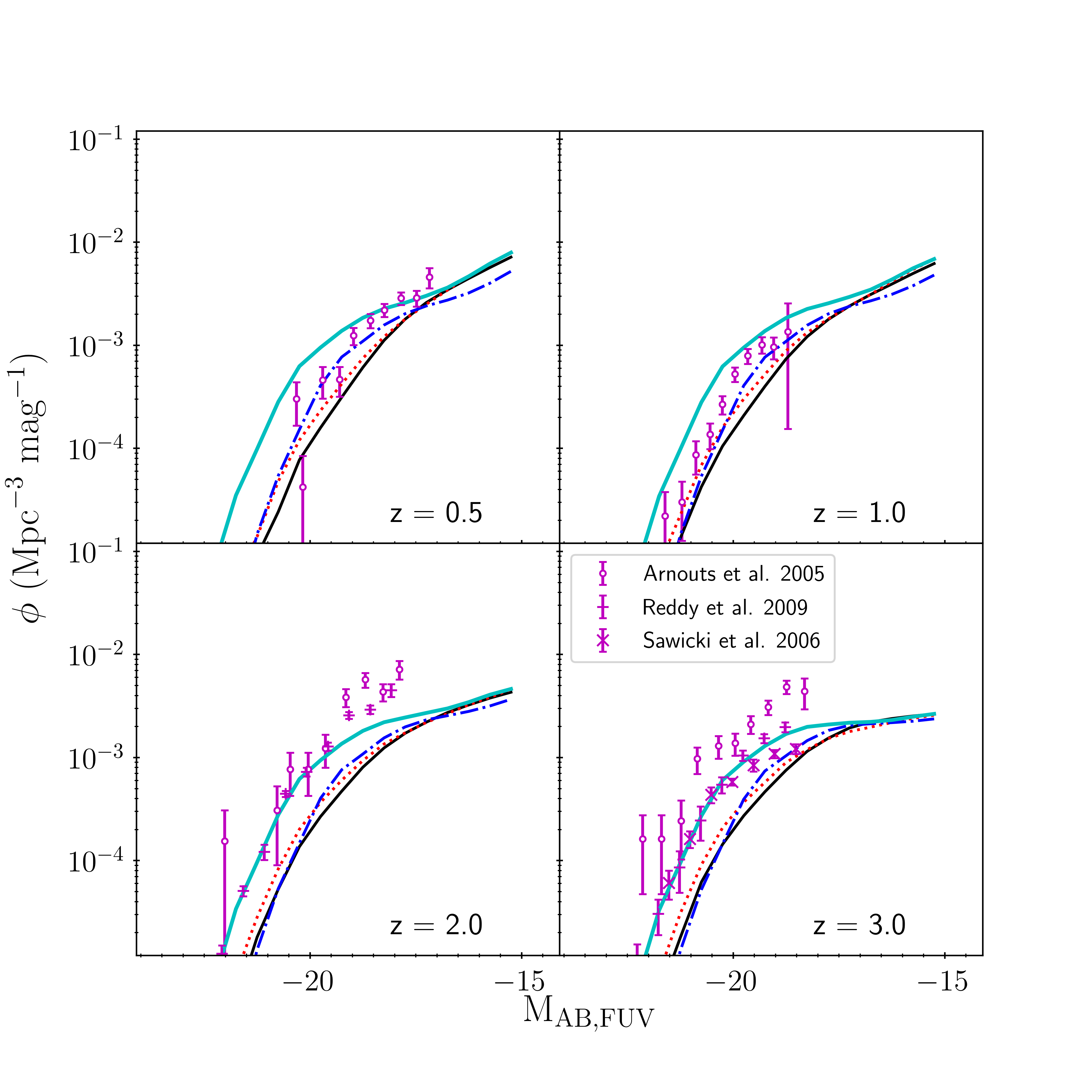}
    \caption{Our predicted far-ultraviolet luminosity functions at $z=0.5, 1.0, 2.0, 3.0$. We show four model variants with different attenuation formula as described in Table \ref{tab:variant}: the cyan, black, red dotted and blue dash dotted lines are our Unattenuated, Default, Somerville and CF00 models, respectively. The symbols with error bars are the observational values from \citet{Arnouts05}, \citet{Reddy09} and \citet{Sawicki06}, as indicated in the legend.}
    \label{fig:UV-evol}
\end{figure}

Figure \ref{fig:UV-evol} shows our predicted far-ultraviolet luminosity functions from $z=0.5$ to $z=3$, compared with the observational results of \citet{Arnouts05, Reddy09} and \citet{Sawicki06}. At $z=0.5$ and $z=1.0$, all of our models with attenuation show reasonable agreement with the observations. In these model variants, our galaxies get fainter in the ultraviolet as redshift increases. This is consistent with the finding of the \textsc{SHARK} SAM \citep{Lagos19} which predicts an increasing attenuation from $z=0$ to $z=3$. 

Our model variants with attenuation (solid black, dashed blue and dotted red lines) systematically underproduce the number of galaxies in the far-ultraviolet at $z=2$ and $z=3$, with a $\sim 0.5$ dex offset. Meanwhile, the Unattenuated variant with pure stellar spectra at these redshift closely follows the observational estimate, showing that the model produces just enough UV luminous galaxies at these redshifts, requiring very little or no attenuation at all. The underproduction of UV luminosity functions grew more significant as the redshift increased, a characteristic also found on previous SAMs \citep{Guo09, Somerville12}.

Observations have found the existence of dust in high redshift galaxies, which obscures the stellar light and heavily influences the far-ultraviolet emission. The mismatch with our predicted luminosity functions at $z=2$ and $z=3$ might be caused by an insufficient early star formation rate, inaccurate dust mass, or the combination of both. The problem of the low star formation rate was encountered by \citet{Guo09} who used a SAM to predict the high redshift galaxy population. The authors found that the ``physical'' SFR from their model at $z=2$ and $z=3$ was significantly lower than the values they derived using dust-corrected UV magnitudes from the same model. The root of the problem was the poor representation of the mean attenuation factor they used to derive the ``mock observed SFR''. 

\dustysage uses a compiled observational SFR from \citet{Somerville01} as a constraint. The compilation is derived from various observations using optical nebular emission lines and the far-UV continuum as the SFR tracer. However, some fraction of star formation activity might also be hidden by dust and can only be observed with infrared tracers. Figure 1 in \citet{Casey18} shows that the obscured SFR from infrared observations is higher than the value derived in UV. The hidden star formation might be the reason for the mismatch between the observed luminosity function and predicted by \mentari. Unfortunately, current infrared data are severely incomplete at $z \approx 2.5$, limiting our understanding of the star formation activity in the early Universe. Due to this incompleteness, we refrain from using the infrared data as a constraint of \dustysage as it will cause inconsistency between our local and high-redshift results.

\begin{figure}
    \centering
    \includegraphics[width=1.1\linewidth]{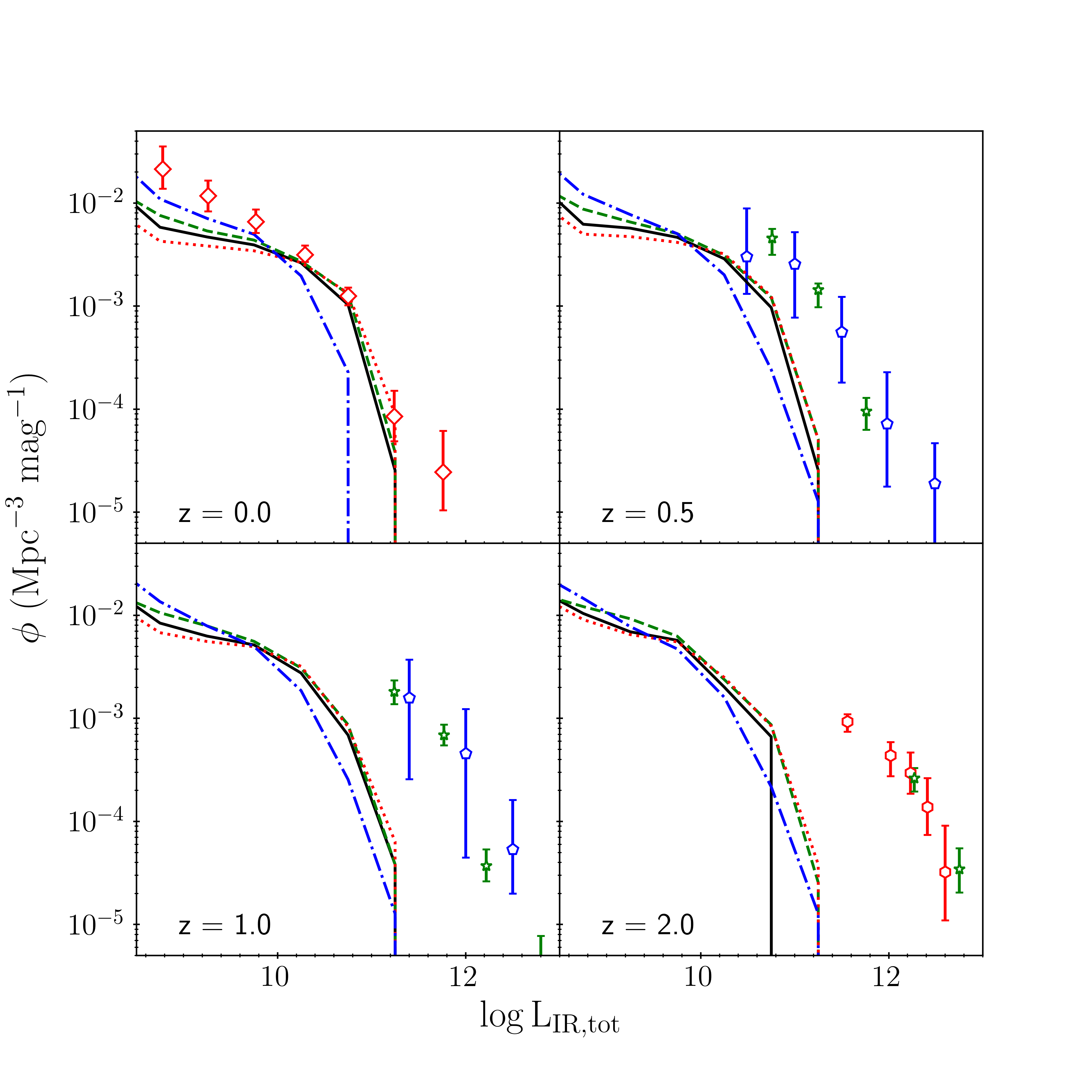}
    \caption{Our predicted luminosity functions for the total infrared emission ($8 - 1000 \mu \mathrm{m}$) at $z=0, 0.5, 1.0, 2.0$. We show four model variants with different attenuation formula as described in Table \ref{tab:variant}: the cyan, black, red dotted and blue dash dotted lines are our Unattenuated, Default, Somerville and CF00 models, respectively. The symbols with error bars are the observational values from \citet{Rodighiero10, LeFloch05}, and \citet{Caputi07} as indicated in the legend. }
    \label{fig:LIR-evol}
\end{figure}

The challenge in the infrared also extends to the measurement of the dust properties of galaxies. Dusty star-forming galaxies are rarely found due to their high obscuration \citep{Whitaker17}, so current far-infrared and submillimetre surveys are biased towards the brightest galaxies \citep{Casey14}. Figure \ref{fig:LIR-evol} shows our prediction for integrated infrared luminosity from $8 - 1000 \mu \mathrm{m}$ at $z = 0, 0.5, 1.0$ and $2.0$. While our default model provides a good match with observation at $z=0$, the discrepancy is clear at higher redshifts. 

Besides SFR, another possible cause for the mismatch is an inaccurate prediction of dust content at high redshift from \dustysage. Figure 3 in \citet{Triani20} shows that the dust mass functions predicted by \dustysage at $z \ge 1$ are systematically lower than that observed. The mismatch of dust mass function at high redshift is found in most galaxy formation models with detailed dust treatments \citep{McKinnon17, Popping17, Vijayan19, Triani20}. We will explore how star formation and dust properties influence the luminosity function in both K-band and far-UV in Section \ref{sssec:parameter-effect} below.

\subsubsection{Effect of \dustysage parameters}
\label{sssec:parameter-effect}

In this section, we explore the processes and parameters in \dustysage sensitive to the UV and K-band luminosity functions, especially at high-redshift. The purpose is not to solely match the high redshift observations but rather to understand what aspects of the physics the observations appear most sensitive to. 

Our prediction for the luminosity function at $z>0$ in Section \ref{ssec:zLF} suggests that the model produces too low a star formation rate and an insufficient dust mass at high redshift. To rectify the SFR problem, we focus on the processes in \dustysage that regulate star formation activity. We find that an influential parameter governing the star formation rate at high redshift is the radio-mode AGN feedback efficiency (RME). The main power source of the radio mode is the accretion of materials onto the supermassive black hole. The energy from the accretion heats the surrounding gas in the halo and suppresses the cold gas supply, leading to the eventual quenching of the star formation activity. The RME dictates how efficiently matter is converted into energy to heat the surrounding gas. Lowering the efficiency results in lower heat, so more gas can cool and be converted into stars, increasing the SFR. 

\begin{figure}
    \includegraphics[width=1.1\linewidth]{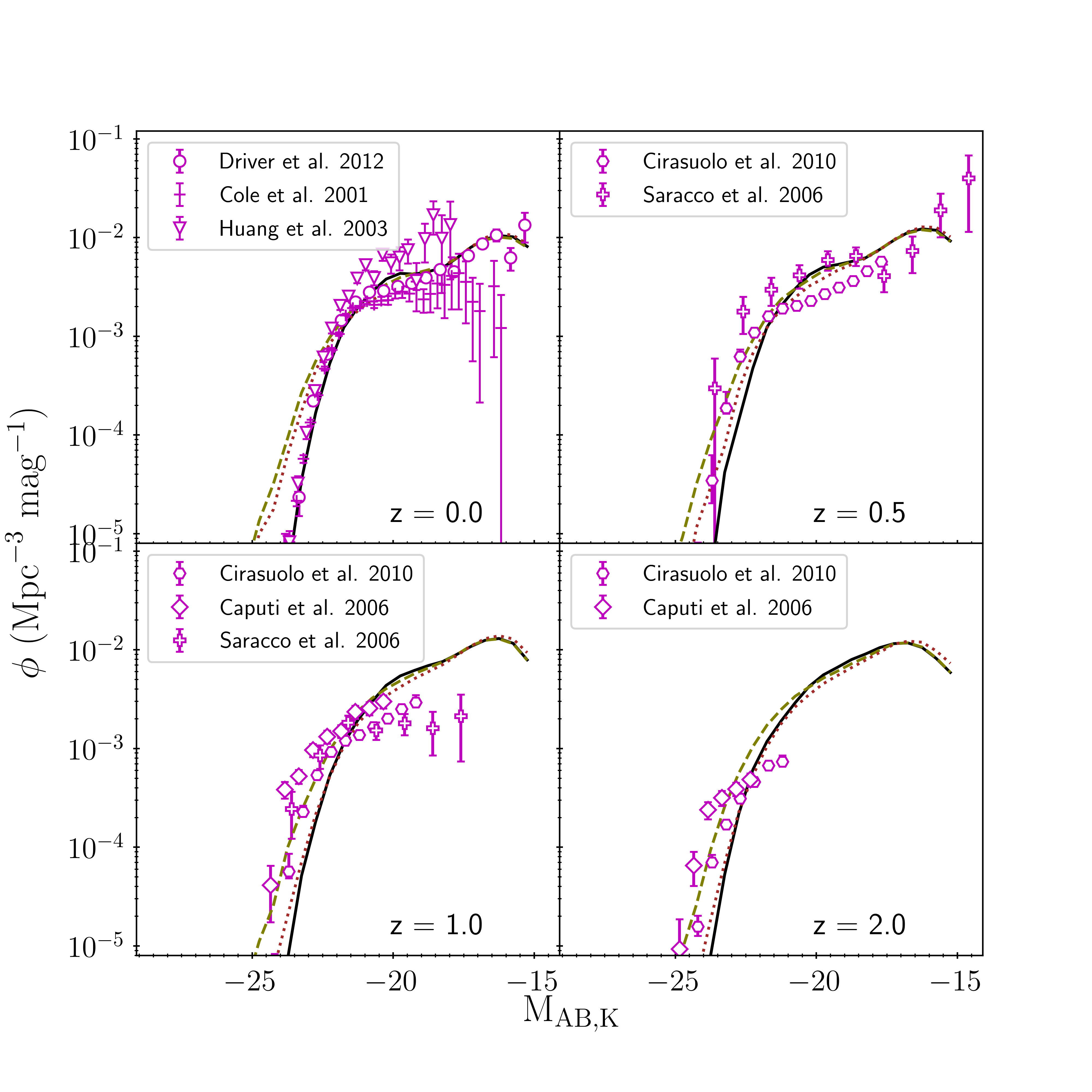}
    \centering
    \caption{Our predicted K-band luminosity functions at $z=0, 0.5, 1.0, 2.0$. Lines with different colour and styles marks different parameterisation of \dustysage. The black, olive dashed, and brown dotted lines show \dustysage with the default parameters, low radio-mode efficiency, and low radio-mode efficiency with no dust destruction, respectively. The symbols with error bars are the observational values from \citet{Cirasuolo10}, \citet{Saracco06}, and \citet{Caputi06}, as indicated in the legend.}
    \label{fig:K-evol-diff}
\end{figure}

\begin{figure}
    \includegraphics[width=1.1\linewidth]{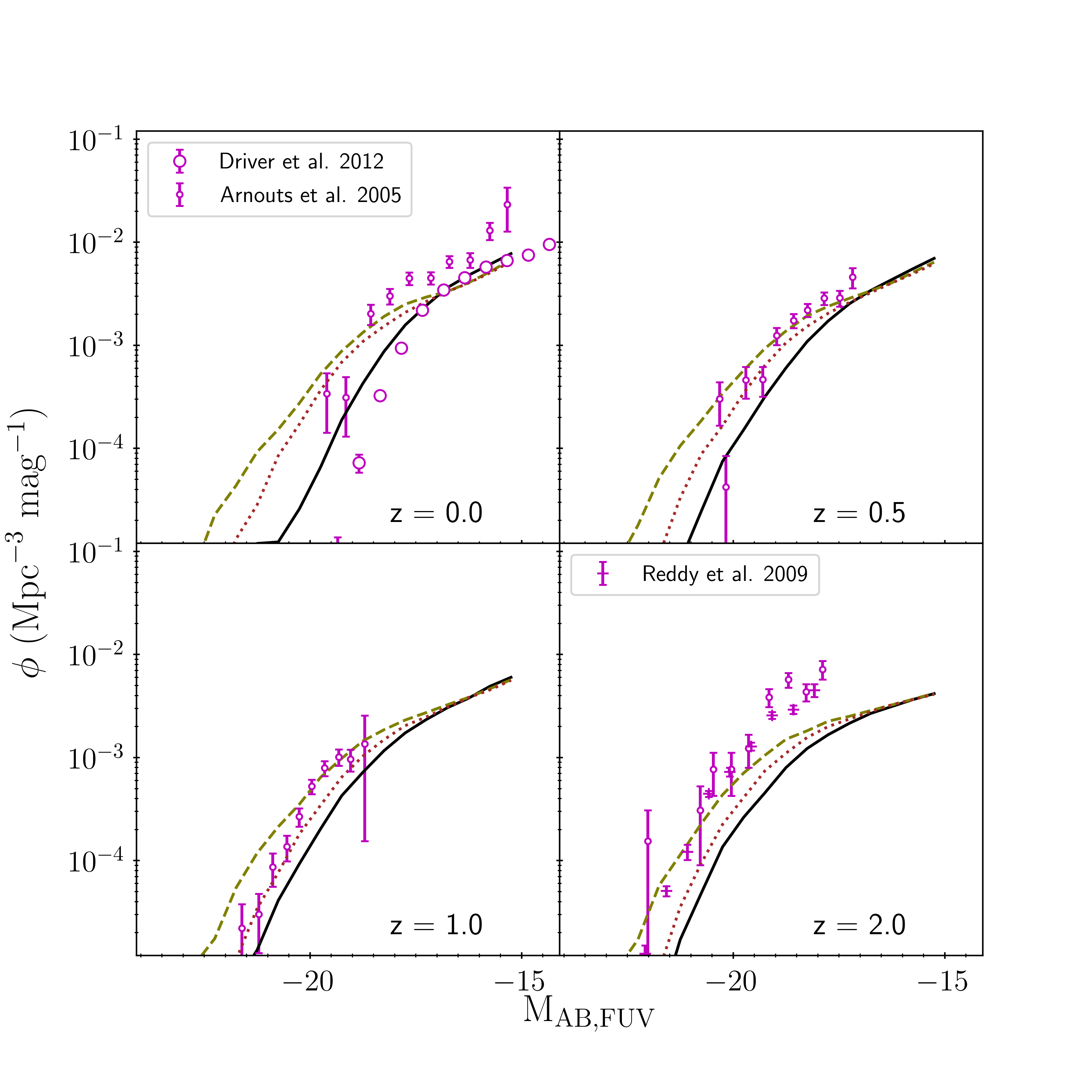}
    \centering
    \caption{Our predicted far-UV luminosity functions at $z=0, 0.5, 1.0, 2.0$. Lines with different colour and styles marks different parameterisation of \dustysage. The black, olive dashed, and brown dotted lines show \dustysage with the default parameters, low radio-mode efficiency, and low radio-mode efficiency with no dust destruction, respectively. The symbols with error bars are the observational values from \citet{Arnouts05}, \citet{Reddy09} and \citet{Sawicki06}, as indicated in the legend.}
    \label{fig:UV-evol-diff}
\end{figure}

The treatment of radio-mode feedback in cosmological models, including \dustysage, is simplistic and it is often unclear how the model relates to the physical feedback mechanisms. This is because current cosmological models do not resolve the scales needed to resolve the accretion of matter onto the black hole and the ejection of an AGN wind. \cite{Schaye15}, for example, found that the behaviour of AGN feedback in the \texttt{EAGLE} simulation changed with resolution, even when all parameters were kept constant. Although it is tricky to compare observed AGN feedback efficiencies with such theoretical prescriptions in galaxy evolution models \citep{Harrison18}, observations of warm outflow gas density suggest that it is far below the $5-10 \%$ range usually adopted in cosmological simulations \citep[e.g.,][]{Santoro20}. In our investigation, we lower the efficiency to $1 \%$ from its fiducial $8 \%$ value. The olive dashed line in Figure \ref{fig:K-evol-diff} and \ref{fig:UV-evol-diff} shows this model with lower efficiency. 

To increase the dust mass at high redshifts in the low RME model, we turn off the destruction via shocks generated by supernovae. The origin of dust, especially in early galaxies, is still an active research topic. Therefore, there is no clear indication on which dust-related mechanisms need to be included in the model. At these epochs, we find this is the simplest way to ensure the increase of dust mass. The brown line in Figure \ref{fig:K-evol-diff} and \ref{fig:UV-evol-diff} presents the model with low RME and no dust destruction. 

Figure \ref{fig:K-evol-diff} shows the redshift evolution of the K-band luminosity function from the model with low RME (olive dashed line) and model with both low RME and no destruction (brown dotted line) compare to the default model (black solid line). Both new model variants overproduse the $z=0$ luminosity function but improve the agreement at higher redshift. At $z=0.5$, the low RME model slightly overproduces the bright end of the luminosity function, indicating the need of more attenuation. The model with no dust destruction, and hence higher dust mass and attenuation, solve this tension. At $z=1$ and $z=2$, the model with low RME provide a reasonable agreement with the observation.

Figure \ref{fig:UV-evol-diff} shows the redshift evolution of the far-UV luminosity function, with lines marking the same model versions as Figure \ref{fig:K-evol-diff}. In the UV, the model with low RME is in reasonable agreement with the observations at $z=2$ but overproduces the number density of galaxies at $z < 2$. This overproduction indicates that we have too much star formation activity that produce UV photons and need more dust to absorb their emission. Our second alternate model with no dust destruction by SN shocks and low RME, is shown as the brown dotted line in Figure \ref{fig:UV-evol-diff}. As expected, this variant is in better agreement with the observation at $z=0.5$ and $z=1$. 

Unfortunately, these new parameter combinations successes do not extend to $z=0$. They struggle to match the $z=0$ observations because of too high of a SFR, which leads to overproduction of the bright end of both the K-band and UV luminosity functions. Hence, the best-fit \dustysage model has different parameter sets at different redshifts. This initial investigation confirms the difficulty of theoretical models with the treatment of AGN feedback and dust formation. It is possible that these feedback and dust processes evolve with redshift, unlike our current prescriptions in \dustysage. Our future work will focus on these aspects. We will explore the interactions of our dust and star formation prescriptions, and other model aspects, across a broader redshift range. These improved prescriptions will allow additional discrimination of the accuracy of the galaxy model and provide new ways to explore the physics and what may be missing at different redshifts.

\section{Discussion}
\label{sec:discussion}

In this paper, we have coupled the stellar and dust properties from the \dustysage SAM to the new generative SED pipeline \mentari. We discuss insights that can be taken from our model predictions in this section.

\subsection{Insight from model variants}
\label{ssec:insights}
 
We have explored several dust attenuation prescriptions and far-infrared templates using \mentari. Our Default model that follows the theoretical scaling of the dust attenuation parameters from the EAGLE simulation \citep{Trayford20} reproduces the observed redshift $z=0$ ultraviolet to near-infrared luminosity functions well. In this variant, we assume that the dust optical depth is correlated tightly with the dust surface density, which is computed directly in \dustysage. This success reflects back to our stellar and dust treatment in \dustysage. The good agreement of our predicted optical to near-infrared emission with the observations indicate the robustness of our star formation and metallicity histories, while the agreement in the ultraviolet to optical windows suggested that our dust prescription in \dustysage has produced sufficient dust to attenuate the stellar emission. The Somerville model variant that uses an empirical scaling \citep{Somerville12} also performs well at reproducing the ultraviolet to near-infrared emission of local galaxies. However, this is expected because the scaling is adjusted to match the ultraviolet luminosity function. Compared to the other variants, the CF00 model that adopts fixed attenuation parameters from \citet{CF00} performs worst at reproducing the ultraviolet luminosity function. This model variant overproduces the number of bright ultraviolet galaxies, which suggests that the attenuation is too low. Note that in this variant, we do not use the dust properties from \dustysage to infer the attenuation parameters. 

We then apply the \citet{Dale14} mid-infrared templates to our Default, Somerville and CF00 model to explore how our model variants behave in the IRAC bands. Emission in these bands are a contribution of the old stellar populations, dust reradiation of the absorbed stellar spectra, and AGN luminosity. In the IRAC $3.6 \mu \mathrm{m}$ and $4.5 \mu \mathrm{m}$ bands where stars are still the dominant contributor of the emission, we find that all models match the observed dataset from \citet{Dai09} remarkably well (see Figure \ref{fig:mir_a}). However, in the $5.8 \mu \mathrm{m}$ and $8.0 \mu \mathrm{m}$ bands where the AGN pollution starts to become significant, our model massively underpredicts the luminosity function. When applying the \citet{Dale14} in our model variants, we keep the AGN fraction as zero. Figure \ref{fig:IR_templates} shows that increasing the AGN fraction parameter ($\mathrm{f_{AGN}}$) in the \citet{Dale14} template can increase the mid-infrared flux and improve our match to the observations. We look forward to properly addressing the AGN factor in the mid-infrared and improving AGN treatment in both \dustysage and \mentari in the near future. 

In the far-infrared, we want to explore how different dust attenuation and infrared templates modify the luminosity functions. First, to determine the effects of dust attenuation in the far-infrared luminosity functions, we apply the \citet{Safarzadeh16} far-infrared templates to the Default, Somerville and CF00 model variants. While both the Default and Somerville models reasonably agree with the observations, the CF00 variant underproduces the far-infrared emission. Their underproduction is caused by insufficient attenuation of stellar light in the ultraviolet to near-infrared window. 

Then, to test how different infrared templates behave in the far-infrared windows, we apply two different templates in the Default model, keeping all other parameters as the same. We compare the \citet{Safarzadeh16} template to the \citet{Dale14} template (Figure \ref{fig:fir_b}). Both templates provide a good agreement with the observations in the $24 \mu \mathrm{m}$, and $160 \mu \mathrm{m}$ range. At $250 \mu \mathrm{m}$, $350 \mu \mathrm{m}$ and $500 \mu \mathrm{m}$ bands, the Dale model provides a better prediction for faint galaxies where the Default model slightly underproduces the number density, while the Default model gives a better prediction for bright galaxies where Dale overproduces the number density. 

In general, our model variants perform well in reproducing the $z=0$ ultraviolet to far-infrared luminosity functions simultaneously. This improves the predictions from many previous SAMs that fail to produce the number of bright submillimetre galaxies \citep{Baugh05, Lacey16, Somerville12}. The main difference between our model and these models is that we vary the attenuation parameters with dust properties computed self-consistently from \dustysage. \citet{Lagos19} using the \textsc{SHARK} SAM also reproduced the ultraviolet to far-infrared galaxy emission by varying the attenuation paramaters. Although they did not directly compute dust mass and used gas phase metallicities and gas mass as proxies for dust instead. 

\subsection{Limitations}
\label{ssec:limitations}

Our model predictions are less successful at reproducing the observations for high redshift galaxies compared to the local population. In this section, we discuss the caveats that limit \mentari and affects its predictive power.

Our default model provides a reasonable fit to the $z=0.5$ and $z=1$ K-band luminosity functions. This match indicates that our galaxies have the correct evolution of stellar properties, including total stellar mass, age and metallicity. However, we slightly underestimate the number density of the brightest galaxies at $z=2$ and $z=3$. There are several suspects for such mismatch. First, it could be caused by the simplistic AGB prescriptions in the BC03 spectral library. Previous studies have found that using a stellar population synthesis model that enhances the near-infrared spectra from AGB stars \citep[e.g.,][]{Maraston05} can resolve this tension \citep{Henriques10, GP14}. However, it could also be caused by the underprediction of the SFR by \dustysage, as discussed in Section \ref{sssec:parameter-effect}.

At high redshift, our Default and Somerville variants produce too significant attenuation and underestimate the observed ultraviolet luminosities (Figure \ref{fig:UV-evol}). Our predicted luminosity functions with no attenuation provide a good fit up to $z=3$, indicating that our total star formation is sufficient to reproduce the ultraviolet emission. However, at $z=2$ and $z=3$, our unattenuated luminosity function is at the borderline of the observed dataset, giving no room for attenuation. Again, this shows the needs to increase the SFR in the model and adjust the attenuation with redshift. However, modifications of attenuation with redshift should be taken with caution as the nature of dust at high redshift remains unclear. For example, observations by \citet{Reddy10} suggest that attenuation decreases with increasing redshift, while \citet{DEE03, Dunne11} find more dusty galaxies in their sample at $z=2$ than in the local Universe.

Changing the parameters in \dustysage to increase the star formation rate at high redshift improves the agreement between our predicted luminosity functions and observation. By lowering the radio-mode efficiency from $8 \%$ to $1 \%$, we match the observed K-band luminosity function in the redshift range $0.5 - 2$. The lower radio-mode feedback efficiency allows stars to form earlier in time so we can reproduce the bright end of the luminosity function where our default model struggles. Consequently, we have too many massive galaxies at $z=0$ and overproduce the number density of local galaxies. The implementation of feedback in the model assumes constant efficiency throughout cosmic time. We possibly need an evolving feedback efficiency with redshift to regulate star formation to match the luminosity and stellar mass function at low and high redshift simultaneously.

The evolution of the UV luminosity function is more complex than the K-band due to the effect of dust attenuation. Here we find we need to adjust the SFR and dust properties in \dustysage to improve the agreement in the far-UV band. Firstly, the UV emission increases as we increase the SFR in the model by dropping the radio-mode efficiency. The prediction from this variant agrees well with the observations at $z=2$ but overpredicts those at $z < 2$. We thus need to increase the dust attenuation in the model to tone down the prediction. Assuming that the relation between dust surface density and dust attenuation is constant across redshift, the simplest solution is to turn off dust destruction to maximise the predicted dust surface density. This results in a good match to the observation at $z=0.5$ and $z=1$. However, even this maximum attenuation is not able to dampen the UV emission to the observed level at $z=0$, a sign that the SFR is too high. Again, this shows the limitation of the model to regulate SFR to match the observations of local and distant galaxies consistently. 

However, caution should be taken when constraining the SFR from observations as a high fraction of star formation is obscured by dust and can be only identified in the infrared \citep{Casey18}. The obscuration level depends on the dust abundance, which evolves with redshift. This challenges our current understanding of dust and metal enrichment processes in the ISM and how they evolve with cosmic time. In the ALMA REBELS survey, \citet{Algera22} find the obscured SFR contributes between $30 \%$ to $60 \%$ at $z \approx 7$.

\subsection{Future application of \mentari}
\label{ssec:mentari_app}

In this work, we have applied our new generative SED pipeline, \mentari, to the \dustysage SAM. Currently, \mentari is open source and available for public use in two modes: the website\footnote{\url{https://share.streamlit.io/dptriani/mentari_web/main}} and full version\footnote{\url{https://github.com/dptriani/mentari}}. At the website, users can generate an SED for a single galaxy by varying stellar and dust parameters. There is an option to input AB magnitudes in various filters so users can compare observed datasets with synthetic fluxes. 

The full version is locally installable and has more features, including constructing the SEDs of multiple galaxies with flexible star formation and metallicity histories, and several options for different attenuation prescriptions and infrared templates. There are hundreds of filters available to convolve the SED into photometries. The full version is useful to create mock catalogues for a large survey, since it allows one to generate the SED of millions of galaxies in one go. The stellar and dust parameters can be input manually or extracted from a galaxy evolution model. Currently, \mentari has specific functions to extract parameters from the \dustysage and \sage SAMs. This function will be expanded in the future to be more general and accept more galaxy evolution models. A fitting function is yet to be included in both website and the full version. Once incorporated, this function will allow users to fit observed datasets directly to a SED built from a galaxy evolution model.

\section{Summary and Conclusions}
\label{sec:conclusion}

This paper has focused on the translation of star formation and stellar light in a galaxy into its panchromatic SED, as would be measured in observations. A critical component of this is the self-consistent modelling of dust and attenuation effects. We present predictions for the local and high-redshift galaxy luminosity functions and galaxies' cosmic spectral energy distribution from the far-ultraviolet to far-infrared. Our predictions are made using the \dustysage SAM \citep{Triani20}, which includes detailed dust tracking on top of the usual modelling of galaxy evolution. The dust processes in \dustysage include stellar dust production, grain growth in molecular clouds, grain destruction by supernova shocks, and thermal sputtering in the halo. \dustysage is one of the first SAMs to use computed dust properties from such complicated processes to infer the galaxy emission. In this work, we run the model on merger trees constructed from the Millennium N-body simulation.

To model the theoretical spectra from galaxies, we introduce the \mentari pipeline, which we run as a post-processing step for \dustysage. \mentari extracts star formation and metallicity histories from \dustysage and combines them with the \citet{2003BC} stellar population synthesis model to produce the intrinsic stellar emission of each galaxy from the far-ultraviolet to near-infrared. We adopt the two-component attenuation model of \citet{CF00} that includes (i) denser dust in birth clouds that envelope young stars and (ii) dust in the diffuse interstellar medium. To compute the \citet{CF00} attenuation parameters, we explore several approaches: (i) adopt the fiducial values of \citet{CF00} as fixed attenuation parameters for all galaxies; (ii) use varying attenuation parameters with dust properties, based on an empirical scaling relation from \citet{Somerville12}; and (iii) similar with the second method, but use the theoretically motivated scaling relations from \citet{Trayford20} and \citet{Lacey16} (following the approach of \citet{Lagos19}). 

The main challenge we grapple with is accounting for dust radiation. We try several different prescriptions. First, we infer the mid-infrared emission from galaxies using the \citet{Dale14} templates. We assume that the total attenuated luminosity equals the total infrared luminosity based on an energy balance principle. Then, we use a correlation between the total infrared luminosity and the $\alpha$ parameter of \citet{Dale14} from \citet{Rieke09} to determine which template to apply to each galaxy. In the far-infrared regime, we explore the \citet{Dale14} templates and a more theoretically based set of templates from \citet{Safarzadeh16}. The latter is derived from the \textsc{SUNRISE} radiative transfer code implemented in a hydrodynamical simulation. The total infrared luminosity and the computed dust mass are used directly to determine which \citet{Safarzadeh16} template to use for a galaxy.

In the ultraviolet wavelengths at $z=0$ (see Figure \ref{fig:UV}), our model variants that use varying attenuation curves give better agreement with the observed far and near-ultraviolet luminosity functions, compared to those with fixed attenuation, which overproduce bright ultraviolet galaxies. 

In the optical and near-infrared, all of our model variants produce a very good agreement with the observed luminosity functions at $z=0$ (Figure \ref{fig:optical}). In this regime, the attenuation effect is less important, as the SED is dominated by stellar light. 

In the mid-infrared, we compare our prediction at $z=0$ with observations in various IRAC bands. These compare well except in the $8.0 \mu \mathrm{m}$ band, where we underestimate the observed luminosity function systematically. The $8 \mu \mathrm{m}$ emission is especially complicated because it includes a significant contribution from unidentified infrared emission (UIE), mainly correlated with PAHs. In addition, the flux at this wavelength is also affected by AGN emission. To compute emission in the mid-infrared, we adopt the templates from \citet{Dale14}. This model may perform better if we implement a realistic AGN fraction parameter instead of fixing it at zero. We plan to implement AGN and focus on the mid-infrared flux in the near future.  

Our model is also successful at reproducing the far-infrared emission at $z=0$ for a wide range of wavebands (Figure \ref{fig:fir_a} and \ref{fig:fir_b}). The CF00 model variant with a fixed attenuation produces less flux across the far-infrared wavelengths compared to the other model variants as a consequence of having less attenuation in the ultraviolet bands (see Figure \ref{fig:UV}). This strengthens our argument that self-consistent dust attenuation is essential for reproducing the full spectrum galaxy emission, from the far-ultraviolet to far-infrared simultaneously.

Finally, we integrate the emission of all local galaxies in our model to compute the cosmic spectral energy distribution at $z=0$ (Figure \ref{fig:csed}). Our results roughly match the observed values from \citet{Andrews17}. The largest discrepancies seen in this comparison are in the mid-infrared, mainly due to PAH emission and AGN contamination, as discussed. We also find a slight mismatch in the model at wavelengths $\ge 500 \mu \mathrm{m}$. A possible cause for this tension is the lack of constraints from the observations of \citet{Andrews17} at the longest wavelengths. 

At high redshift, our model underestimates the observed UV luminosities (Figure \ref{fig:UV-evol}). Modifications to the radio-mode efficiency and dust destruction process in \dustysage improves the agreement but is no longer consistent with the $z=0$ observations (Figure \ref{fig:UV-evol-diff}). In Figure \ref{fig:K-evol}, we show that our model fits well the observed K-band luminosity function up to $z=3$. Its excellent match indicates that we are producing the correct evolution of stellar properties, including the total stellar mass, age and metallicity. However, at $z=2$ and $z=3$, we slightly underestimate the number of the brightest galaxies. Increasing the SFR by modifying the radio-mode efficiency solves this problem (Figure \ref{fig:K-evol-diff}). But again, this change creates tension with the $z=0$ observations. Further study is needed to match at $z=0$ and $z > 0$ simultaneously. Previous work has found that using a stellar population synthesis model that enhances the near-infrared spectra from AGB stars \citep[e.g.,][]{Maraston05} can also alleviate this tension \citep{Henriques10, GP14}.

In this paper, we have presented a comprehensive framework to produce the panchromatic galaxy emission from a SAM with a state-of-the-art dust model. We find that a detailed dust treatment is necessary to generate realistic observables, both for local and high-redshift galaxies. The strong agreement of our predictions with an extensive set of observations reflects how the model accurately describes various complicated processes in galaxy evolution and its discrepancies provide opportunity to explore the complex interactions of galaxy physics and dust, and how these translate into galaxy light. Our pipeline with \dustysage and \mentari produces theoretical predictions to compare to future observations using JWST and ALMA, which will constrain dust emission and properties in early galaxies at increasing level of detail.

\section*{Acknowledgements}

DPT thanks Haru Hawaari for carrying the study on \dustysage parameters and its effect on stellar mass function, Harry Ferguson for providing suggestion on the infrared templates, Elisabete da Cunha and Carlton Baugh for insightful discussion and comments on the thesis chapter which is the basis of this paper. This research was supported by the Australian Research Council Centre of Excellence for All Sky Astro-physics in 3 Dimensions (ASTRO 3D), through project number CE170100013. CP is supported by the Canadian Space Agency under a contract with NRC Herzberg Astronomy and Astrophysics. The Semi-Analytic Galaxy Evolution (SAGE) model, on which \dustysage was built, is a publicly available codebase that runs on the dark matter halo trees of a cosmological N-body simulation. It is available for download at \url{https://github.com/darrencroton/sage}. This research has used \texttt{python} (\url{https://www.python.org/}), \texttt{numpy} \citep{Vanderwalt11} and \texttt{matplotlib} \citep{Hunter07}.

\section*{Data Availability}
The data underlying this article are available in the article. The galaxy formation model used to generate the data is available at \url{https://github.com/dptriani/dusty-sage}. The pipeline used to generate SED in this article is available at \url{https://github.com/dptriani/mentari}.



\bibliographystyle{mnras}
\bibliography{bibliography} 





\bsp	
\label{lastpage}
\end{document}